\newif\ifanonymized
\anonymizedfalse    

\newif\ifsubmit
\submittrue	

\documentclass[acmlarge]{acmart}
\usepackage[acronym]{glossaries}
\makeglossaries
\newacronym{gloss}{GLOSS}{Group of LLMs for Open-ended Sensemaking}

\usepackage{fancyhdr}
\usepackage{xcolor}
\usepackage{colortbl}

\usepackage{url}

\usepackage[mmddyyyy,hhmmss]{datetime}

\usepackage{xspace}
\usepackage{soul}
\usepackage{subcaption}
\usepackage{graphicx}
\usepackage{tabularx}
\usepackage{makecell}

\usepackage{booktabs}
\usepackage{placeins}

\definecolor{customblue}{HTML}{41cffa}
\usepackage{graphicx}
\DeclareGraphicsExtensions{.pdf,.png,.eps,.ps,.jpg}
\newcommand{\filleddot}{\tikz\draw[scale=0.3,draw=black,fill=customblue] (0,0) circle(0.3cm);}
\newcommand{\emptypdot}{\tikz\draw[scale=0.3,draw=black,fill=white] (0,0) circle(0.3cm);} 
\newcommand{\halfFilledDot}{%
    \tikz[scale=0.3]{
        \begin{scope}
            \clip (0,0) circle (0.3cm); 
            \fill[customblue] (-0.3cm,-0.3cm) rectangle (0,0.3cm); 
        \end{scope}
        \draw[black] (0,0) circle (0.3cm);
    }
}

\usepackage{tikz}

\definecolor{rowcolor1}{RGB}{200, 236, 247}
\definecolor{rowcolor2}{RGB}{237, 237, 237} 

\newcommand{\circlednumber}[1]{%
  \tikz[baseline=(char.base)]{
    \node[shape=circle, draw=black, inner sep=1pt, minimum size=10pt] 
    (char) {\scriptsize #1};}}

\ifsubmit
  \newcommand{\hey}[1]{\relax}
  \newcommand{\heyvarun}[1]{\relax}
  \newcommand{\bibtex}[1]{\relax}
  \newcommand{\note}[1]{\relax}
\else
  \newcommand{\hey}[1]{\textcolor{magenta}{[{#1}]}}
  \newcommand{\heyvarun}[1]{\textcolor{blue}{[Varun: {#1}]}}
  \newcommand{\note}[1]{\par\textcolor{magenta}{Note: {#1}}\par}
  \newcommand{\bibtex}[1]{\textcolor{red}{@bibtex}\{#1\}} 
\fi

\newcommand{\hide}[1]{\relax}

\newcommand{\seclabel}[1]{\label{sec:#1}}

\definecolor{maroon}{HTML}{800000}

\definecolor{darkred}{HTML}{FF0000}
\newif\ifredtext
\newcommand{\rnr}[1]{%
  \ifredtext
    {\color{darkred}{#1}}%
  \else
    {#1}%
  \fi
}
\usepackage{tcolorbox}
\usepackage{listings}
\usepackage{verbatim}
\usepackage{wrapfig}

\definecolor{customlime}{HTML}{CDE741}

\newcommand{\circlelabel}[1]{%
  \tikz[baseline=(char.base)]{
    \node[shape=circle, draw=black, fill=customlime, inner sep=0.5pt, minimum size=1.1em] (char) {\scriptsize\textbf{#1}};
  }\xspace
}

\redtextfalse

\begin{document}

\title{GLOSS: Group of LLMs for Open-Ended Sensemaking of Passive Sensing Data for Health and Wellbeing}
\date{}

\ifanonymized
    \author{Anonymized for blind submission}
\else
    \author{Akshat Choube}
        \email{choube.a@northeastern.edu}
        \affiliation{%
          \institution{Northeastern University}
          \country{Boston, United States}
        }
        
        \author{Ha Le}
        \email{le.ha1@northeastern.edu>}
        \affiliation{%
          \institution{Northeastern University}
          \country{Boston, United States}
        }

          \author{Jiachen Li}
        \email{li.jiachen4@northeastern.edu}
        \affiliation{%
          \institution{Northeastern University}
          \country{Boston, United States}
        }

         \author{Kaixin Ji}
        \email{li.jiachen4@northeastern.edu}
        \affiliation{%
          \institution{RMIT University}
          \country{Melbourne, Australia}
        }

         \author{Vedant Das Swain}
     \email{v.dasswain@northeastern.edu}
        \affiliation{%
          \institution{Northeastern University}
          \country{Boston, United States}
        }

     \author{Varun Mishra}
     \email{v.mishra@northeastern.edu}
        \affiliation{%
          \institution{Northeastern University}
          \country{Boston, United States}
        }
\fi




\begin{abstract}
The ubiquitous presence of smartphones and wearables has enabled researchers to build prediction and detection models for various health and behavior outcomes using passive sensing data from these devices. Achieving a high-level, holistic understanding of an individual's behavior and context, however, remains a significant challenge. Due to the nature of the passive sensing data, \textit{sensemaking} --- the process of interpreting and extracting insights -- requires both domain knowledge and technical expertise, creating barriers for different stakeholders. Existing systems designed to support sensemaking are not open-ended or cannot perform complex data triangulation. In this paper, we present a novel sensemaking system, \textit{Group of LLMs for Open-ended Sensemaking (GLOSS)}, for open-ended sensemaking capable of performing complex multimodal triangulation to derive insights. We demonstrate that GLOSS \rnr{significantly} outperforms commonly used Retrieval-Augmented Generation (RAG) technique, \rnr{achieving \textit{87.93\%} {accuracy} and \textit{66.19\%} {consistency} compared to RAG's \textit{29.31}\% accuracy and \textit{52.85}\% consistency}. Furthermore, we showcase the promise of GLOSS using four use cases inspired by prior and ongoing work in UbiComp and HCI communities. Finally, we discuss the potential of GLOSS, the broader implications, and the limitations of our work. 
\end{abstract}
\maketitle  



\ifsubmit
    \relax
\else
    \par\noindent \textcolor{red}{\textbf{DRAFT}: \today\ -- \currenttime}
    \pagestyle{fancy}
    \lhead{DRAFT in preparation}
    \rhead{version: \today\ -- \currenttime}
    \chead{}
    \lfoot{}
    \rfoot{}
\fi


\section{Introduction} 
\seclabel{introduction}

Humans continuously process vast amounts of information, constructing explanations to better understand situations or tasks at hand. This process is called \textit{``sensemaking''}, a term widely used by researchers across various disciplines, including Human-Computer Interaction~\cite{pirolli2005sensemaking, russell1993cost}, Organizational Studies ~\cite{greenberg1995blue}, Information Science ~\cite{norman1975role, dervin1992mind}, and others ~\cite{dervin1983overview, barrett2009interpretation, krizan1999intelligence, klein2007data}. It consists primarily of two iterative processes: (1) seeking, extracting, and filtering information, often referred to as the \textit{information-seeking loop} (or the \textit{foraging loop}), and (2) iteratively developing and refining an understanding of the situation, known as the \textit{sensemaking loop}~\cite{zhang2014towards}. Sensemaking is triggered by a \textit{need for information} of a sensemaker (the individual or agent engaged in sensemaking) often expressed through a query. The information-seeking and sensemaking process continues until the sensemaker is satisfied with the developed understanding or explanations. Sensemaking is an innate process individuals engage in regularly -- whether in making sense of conversations, interpreting trends in data, or solving problems. However, when the complexity of the data or the context exceeds an individual's expertise, external help---such as collaboration with others, tools, and frameworks---can enhance their ability to process and understand the information effectively. Data related to health, behavior, and well-being is often complex to interpret due to variability, multidimensionality, and sensitivity to context \cite{arora2008frustrated, clarke2016health}. Therefore, in recent years, researchers have made significant efforts to develop support systems \cite{ramjee2024cataractbot, chikersal2020understanding, qu2008building} and tools \cite{merrill2024transforming, fang2024physiollm, li2024vital} to assist in the sensemaking of health and behavioral data. 


Passive sensing data collected using mobile phones and wearables offer a valuable source of behavioral and health information ~\cite{abdullah2018sensing, mohr2017personal}. Passive mobile sensing has demonstrated significant potential in monitoring and assessing various health and well-being outcomes, including depression ~\cite{xu2023globem, de2013predicting, wang2014studentlife}, stress ~\cite{mishra:context, sano2013stress, mishra:continuous-stress, mishra_evaluating_2020, wang2014studentlife, DasSwain2022semantic}, and anixety \cite{tlachac_deprest-cat_2022}. 
Sensemaking of passive mobile sensing data can offer valuable insights to a broad range of stakeholders, including HCI researchers, behavioral scientists, clinicians, and individuals who actively engage in personal health tracking and informatics. Through sensemaking, HCI researchers and behavioral scientists can understand patterns in human behavior and can design effective interventions~\cite{wu2024mindshift, li2024vital}. Similarly, clinicians can make informed decisions and improve patient care through data-driven insights \cite{adler2024beyond}. Likewise, sensemaking can assist self-health trackers in gaining a deeper understanding of personal well-being ~\cite{cocskun2023data, raj2019clinical}, supporting disease management~\cite{mamykina2017personal}, and improving reflection~\cite{stromel2024narrating, neupane2024momentary}.

Researchers have been using passive sensing to build predictive models for specific behavioral outcomes (such as depression, stress, and physical activity), but a holistic high-level understanding of an individual's well-being remains limited and challenging~\cite{adler2024beyond}.  A key hurdle is the high dimensionality and data missingness of passive sensing data, particularly in its raw form, which makes it difficult to interpret~\cite{choube_imputation_2024}. Furthermore, working with this type of data demands a specialized skill set, including expertise in handling and analyzing multimodal sensor data as well as proficiency in programming languages~\cite{Karahano2024designing}. 
Drawing meaningful insights from passive sensing data often involves triangulating multiple sensor streams. For example, researchers might triangulate several passively sensed data streams, such as location, physical activities, and phone usage, to understand whether a participant had a productive day at work. People with limited coding skills and domain knowledge find it extremely difficult to interpret sensor data, let alone derive meaningful insights from it~\cite{raj2019clinical, choe2014understanding}. Behavioral scientists and clinicians, who often have limited expertise in programming and analyzing sensor data, rely on domain experts to create dashboards and visualizations tailored to their needs. These dashboards, however, often capture only a limited subset of the dimensions of interest. Consequently, to obtain additional data or analysis, these stakeholders may need to seek further assistance from domain experts, making the process time-consuming and effort-intensive~\cite{xu2021understanding, Karahano2024designing}. Self-health trackers face similar challenges, as they are constrained by the predefined visualizations and insights provided by tracking systems/applications, often expressing a desire for deeper or more personalized information~\cite{attig2020abandonment, cocskun2023data, lazar2015we}.
These barriers faced by various stakeholders in sensemaking and deriving insights from passive sensing data highlight the need for an open-ended, user-friendly, and accessible sensemaking system.

Current sensemaking and data exploration tools for passive sensing data are either not open-ended \cite{merrill2024transforming, cosentino_towards_2024} or often deal with simplified tasks like reporting statistical metrics of pre-calculated measures \cite{fang2024physiollm}. Stakeholders require a certain level of domain knowledge and expertise to effectively interact with and use these tools \cite{li2024vital, chen_sensor2text_2024}. For example, many existing systems work with aggregated or processed data \cite{merrill2024transforming}, rather than directly using raw data from sensing streams, necessitating expertise in sensor data preprocessing. Additionally, current sensemaking systems often fall short in delivering customized information tailored to the varying needs of different stakeholders \cite{canali2022challenges}. For instance, most sensemaking systems reduce individuals to quantified beings, failing to provide a holistic view of their health and wellbeing ~\cite{stromel2024narrating}. In recent years, Large Language Models (LLMs) have demonstrated impressive performance in ``common sense'' reasoning~\cite{madaan2022language, davis2023benchmarks}, medical question answering~\cite{singhal2023towards, hua2024large}, and programming tasks~\cite{li2023starcoder, lozhkov2024starcoder}, making them a promising component in the design of sensemaking systems for health and behavioral data ~\cite{merrill2024transforming, fang2024physiollm, stromel2024narrating}.

In this work, we introduce \acrshort{gloss} (\acrlong{gloss}), a novel sensemaking system designed for passive sensing data collected via phones and wearable devices. \acrshort{gloss} uses a network of large language models that collaborate through rules and processes inspired by models of sensemaking in HCI~\cite{pirolli2005sensemaking} and Organizational Studies~\cite{klein2007data}. This fully automated system integrates function calling and code generation capabilities to retrieve and analyze passive sensing data. \acrshort{gloss} is a query-based system where users\footnote{Here, users could be to both humans or automated systems that need to interpret data through querying GLOSS. Throughout the rest of the paper, when we refer to ``users,'' we include both.} can seamlessly express their information needs by posing queries in natural language. \acrshort{gloss} offers a scalable, automated, and off-the-shelf solution, eliminating the need for domain expertise or coding skills and making it highly accessible to diverse stakeholders.
GLOSS can answer simple data retrieval queries like \textit{``What was the last social media app this person used before 11 am?''} to more complex queries like \textit{``Were there anomalies in how this person used their social media''} along with reasoning questions like \textit{``Why did this person self-report a high-stress event on Monday evening?''}. We present an evaluation of the GLOSS's underlying system on objective and subjective queries and compare it against a Retrieval Augmented Generation (RAG) technique. Following this, we demonstrate how GLOSS can drive a variety of outcomes and research methods: a chat-based system for researchers to query \& analyze data, generating personal informatics narratives to support reflection, and even automated tasks like explaining anomalies and personalizing EMA prompts. We keep detailed evaluations for each of the use cases in future work.




Thus, we make the following \textbf{contributions}:
\begin{itemize}
    \item We introduce \acrshort{gloss}, a novel query-based sensemaking system that is open-ended, scalable, and extensible, designed to support diverse stakeholders in interpreting passive sensing data.
    \item We conducted technical evaluations benchmarking the performance of \acrshort{gloss}, demonstrating promising results compared to the RAG technique.
    \item We present the application of GLOSS across four important use cases inspired by previous and ongoing work in the UbiComp and HCI communities.
\end{itemize}

\acrshort{gloss} is the foundational step towards a fully open-ended and multimodal sensemaking system designed to work directly with raw sensor data, paving the way for numerous future applications. Our code for the GLOSS system and results are publicly available on [GitHub]\footnote{https://github.com/UbiWell/GLOSS} to enable other researchers to experiment with GLOSS on their own data and extend this work.






    

\section{Related Work} 
Our work builds on prior research on passive sensing data from smartphones and wearables, the sensemaking process, and recent LLM-driven data exploration systems.

\subsection{Barriers in Interpreting Passive Sensing Data}
Mobile and wearable devices can collect large amounts of passive sensing data on a daily basis. Researchers leverage these longitudinal streams of passive sensing data to predict a variety of health and behavioral outcomes~\cite{bloomfield_detecting_nodate, jacobson_deep_2021, schmidt_introducing_nodate, kanade_activity_2004}. Researchers can use these predictive models to drive context-aware interactive systems, just-in-time interventions, or health monitoring tools\hey{cite}. Although there have been multiple breakthrough attempts to incorporate mobile passive sensing data into human behavioral detection and personal health informatics (PHI) systems \cite{mishra2021detecting, lukic_breathing_2022}, interpreting longitudinal passive data remains challenging for several reasons. First, processing and interpreting passive sensing data often requires domain knowledge and technical proficiency~\cite{xu2021understanding}. Many researchers from non-technical backgrounds, like psychology or behavioral science, might struggle when processing raw sensor data~\cite{felix2019mobile}. Second, the patterns of passive sensing data are often highly individualized \cite{xu2022globem} -- different participants have different behavioral patterns, thus often requiring researchers to train personalized models for each participant~\cite{xu_leveraging_2021, han_systematic_2024}. Different datasets also have varying formats and sensor streams available, often with significant amounts of missing data \cite{engels_imputation_2003, jaques_multimodal_nodate, choube_imputation_2024}. This inconsistency prevents researchers from developing generalized, off-the-shelf models that can be used without any additional training or fine-tuning \cite{habermacher_describing_2020}. Finally, most behavioral detection and PHI focus on a limited predefined health aspects (e.g., depression \cite{lu_joint_2018, xu_leveraging_2021}, anxiety~\cite{tlachac_deprest-cat_2022}, physical activity counts~\cite{botros2016go}), which limits the ability of researchers to ask open-ended questions or triangulate information across different data sources \hey{cite}. A system capable of responding to open-ended, natural language input allows researchers to explore more complex questions and gather insights that might not be immediately accessible in structured formats or require reasoning based on common sense knowledge \cite{wang_towards_2022, jiang_dt2vis_nodate, lundgard_accessible_2022}. These barriers are exacerbated for people interested in understanding personal health behavior through self-tracking. Attig et al. \cite{attig2020abandonment}, found that data incomprehensibility was one of the important reasons of wearable activity tracking attrition. Likewise, Lazar et al. \cite{lazar2015we} argued that being able to understand data from smart devices increases motivation to keep using those devices.
Thus, to assist in interpreting and understanding passive sensing data (sensemaking), researchers have built support systems and tools.

\seclabel{background}

\subsection{Sensemaking of Passive Sensing Data}

\textit{Sensemaking} as a concept gained popularity starting in the late 1970s~\cite{dervin1983overview, norman1975role}. Researchers have expanded on this concept and applied it across various fields, such as organizational theory, education, decision-making, and human-computer interaction~\cite{dervin1992mind, greenberg1995blue, klein2007data, pirolli2005sensemaking}. The broad idea of sensemaking being an iterative process that involves information seeking, and making sense of information remains consistent across these fields~\cite{zhang2014towards}. 
Sensemaking of passive sensing data is a complex and challenging task. In prior work, researchers have developed support systems and networks to share passive sensing data with others, facilitating discussions about their health. Puussaar et al. built a social network for self-trackers \cite{puussaar_enhancing_2017}, while Coşkun suggested building social communities and groups for collaborative sensemaking~\cite{cocskun2023data}. While these approaches can assist in sensemaking, they often rely on the expertise of other people and are often not widely accessible. 

Researchers have also explored innovative data visualization techniques to support users with the interpretation of passive sensing data. These approaches, however, do not take the additional step of performing the sensemaking process for the users. Choe~et~al. introduced a web-based visualization dashboard to support self-reflection and allowing users gather insights on their personal data, but they found that data interpretation was a key hurdle~\cite{choe2014understanding}. Adler~et al., through an in-lab study with clinicians, showed that while visualization dashboards provide actionable insights for treatment, they often lack context and introduce personal biases. Mamykina et al. demonstrated that visualizations aimed at disease management for diabetes patients assisted in sensemaking but were considered complex to interpret due to the need to triangulate data across visualizations~\cite{mamykina2017personal}. Likewise, Karahanoğlu et al. showed that self-health trackers get overwhelmed with the data presented to them if not properly curated \cite{Karahano2024designing}. 

Thus, visualizations-only approaches for passive sensing data are still considered hard to interpret by clinicians, behavioral scientists, self-health trackers, and even researchers. With the advent of Large Language Models (LLM), researchers have started using LLMs' capabilities in common sense reasoning, question answering, and code generation to design systems that can do sensemaking on behalf of the users.

\subsection{LLMs with Passive Sensing Data}

Previous research has used LLMs to develop predictive models for various health and behavioral outcomes, including depression, stress, sleep quality, dementia diagnosis, and activity recognition \cite{kim_health-llm_2024, llmsense, ji_hargpt_2024}. As LLMs primarily process natural language data, these studies convert passive sensing data into textual representations.  Researchers subsequently vectorize these textual representations, storing them in vector databases and employing Retrieval-Augmented Generation (RAG) techniques to retrieve relevant data to queries \cite{lewis_retrieval-augmented_2020, yang_retrieval-augmented_2025}. Yang et al. introduced an AI doctor capable of retrieving sensor data and medical knowledge through RAG  to suggest medical advice \cite{yang_drhouse_2024}. Furthermore, in some studies, researchers have fine-tuned LLMs using wearable data for tasks like sleep quality prediction \cite{cosentino_towards_2024} and question-answering based on activities \cite{chen_sensor2text_2024}. Researchers have also built systems that can analyze passive sensing data and offer insights: Stromel et al. created LLM-based narratives of daily step count data to trigger self-reflective practices \cite{stromel2024narrating}. Likewise, Cosentino et al. \cite{cosentino_towards_2024} fine-tuned a large Medical LLM with sensor data to generate sleep and physical activity insights. Xiong et al. presented a tutorial on building context-aware LLM-based systems to command sensor actuators ( e.g., a mobile z-arm in an apartment for assisted living) that users can operate via natural language commands~\cite{xiong_natural_2023}. In a complementary work to building such LLM-based systems, Li et al. \cite{li2024vital} developed a prototype system called Vital Insight and investigated its' applicability in assisting domain experts' sensemaking process. Some researchers have also argued most existing PHI systems do not offer personalized insights to individuals \cite{rooksby2014personal}, often reducing them to ``quantified beings'' \cite{stromel2024narrating}.

Two closely related works are those by Fang et al. \cite{fang2024physiollm} and Merrill et al. \cite{merrill2024transforming}.
Fang et al. presented an interactive system, PhysioLLM, that is capable of providing insights derived from physiological data. These insights, however, are limited to finding correlations between aggregated data (for instance, between hourly step counts and hourly heart rate) through precomputed correlation matrices. Merill et al. developed PHIA, an LLM-based system that answers user queries through code generation and Google search. PHIA follows a structured three-step (thought-act-observe) process to interpret data but is restricted to working with daily aggregated data stored in CSV files (e.g., daily total steps, daily wake-up time, etc.). The authors evaluated PHIA using template-based queries, thus restricting its adaptability to more open-ended inquiries. Additionally, PHIA’s code generation is confined to basic statistical operations (e.g., mean, median, standard deviation, and correlations), making it less suited for answering complex queries that require deeper data integration and triangulation. We summarize the limitations of existing systems in Table \ref{tab:system_limitations}.

Our proposed system, GLOSS, addresses these limitations: GLOSS retrieves, processes, and analyzes raw passive sensing data directly from databases rather than relying on pre-aggregated or pre-computed data and metrics. Additionally, GLOSS can generate complex code that goes beyond basic statistical operations, enabling integration across multiple data streams and facilitating data triangulation. Moreover, GLOSS does not require any model training or fine-tuning and can easily be extended to include various sensor streams. Thus, we believe GLOSS is the first fully open-ended sensemaking system that enables different stakeholders to analyze and interpret passive sensing data more effectively.  We highlight these differences between prior works and GLOSS in Table \ref{tab:comparison}.

\begin{table}[h!]
\centering
\small
\renewcommand{\arraystretch}{1.2} 
\caption{Limitations of current systems, their descriptions, and example queries that those systems would fail to answer.}
\begin{tabularx}{\textwidth}{>{\raggedright\arraybackslash}p{0.2\textwidth}|p{0.35\textwidth}|p{0.35\textwidth}}
\hline
\textit{Limitation} & \textit{Description} & \textit{Example Query} \\ \hline \hline
Uses Aggregated Data & Some systems rely on data aggregated over hours or days, limiting their ability to provide fine-grained insights. & \textit{What app was the user using around 8:30 pm yesterday? Was it a social media app}? \\
Limited Computation & Existing systems often generate code for simple statistical functions or use precomputed results, limiting their ability to perform advanced calculations. & \textit{Did the user remain within one mile of their home throughout the day on 28th October 2024?} \\
Single Data Stream & Existing systems use a single or limited set of data streams, reducing the insights they can provide. & \textit{---} \\
No Data Triangulation & Some systems use multiple sensor streams but lack the ability to triangulate data between them. & \textit{What apps was the user using when their heart rate spiked over 120 bpm on the 17th of September?} \\
Requires Model Training & Some systems require model fine-tuning, limiting their adaptability and restricting their usability to specific domains. & \textit{---}\\
Impersonal Insights & Many existing systems cannot tailor their responses to the specific needs of the users. & \textit{Summarize the physical activity of the user on 25th August 2024, focusing on qualitative aspects.} \\
Requires Expertise& Some existing systems assist in sensemaking but still need passive sensing or programming expertise. &\textit{---} \\
\hline
\end{tabularx}
\label{tab:system_limitations}
\end{table}

\begin{table}[h]
\small
    \centering
    \caption{A summary of the capabilities of related LLM-based sensemaking approaches, where \protect\filleddot \hspace{0.1cm}indicates the presence of a particular capability in the given approach. \protect\halfFilledDot in Multi data streams denotes the system uses multiple data streams but does not do complex data triangulations}
    \resizebox{\textwidth}{!}{
    \begin{tabular}{l c c c c c c c}
        \toprule
        \textbf{Model} & \makecell{\textbf{Open-} \\ \textbf{ended}} & \makecell{\textbf{Raw passive} \\ \textbf{sensing data}} & \makecell{\textbf{Multi} \\ \textbf{data streams}} & \makecell{\textbf{Query-based} \\ \textbf{interface}} & \makecell{\textbf{No model} \\ \textbf{training}} & \makecell{\textbf{Code} \\ \textbf{generation}} & \makecell{\textbf{Open} \\ \textbf{source}} \\
        \midrule \hline
        Narrating Fitness \cite{stromel2024narrating} & \emptypdot & \filleddot & \emptypdot & \emptypdot & \filleddot & \emptypdot & \emptypdot \\
        PH-LLM \cite{cosentino_towards_2024} & \emptypdot & \filleddot & \halfFilledDot & \filleddot & \emptypdot & \emptypdot & \emptypdot \\
        Vital Insight \cite{li2024vital} & \emptypdot & \filleddot & \halfFilledDot & \emptypdot & \filleddot & \emptypdot & \emptypdot \\
        PhysioLLM \cite{fang2024physiollm} & \emptypdot & \emptypdot & \halfFilledDot & \filleddot & \filleddot & \emptypdot & \emptypdot \\
        PHIA \cite{merrill2024transforming} & \emptypdot & \emptypdot & \halfFilledDot & \filleddot & \emptypdot & \filleddot & \emptypdot \\\hline 
        \textbf{GLOSS (ours)} & \filleddot & \filleddot & \filleddot & \filleddot & \filleddot & \filleddot & \filleddot \\
        \bottomrule
    \end{tabular}
    }
    \label{tab:comparison}
\end{table}

\section{System Design}


GLOSS consists of a network of Large Language Models (LLM) designed to simulate the process of sensemaking in humans, drawing inspiration from the human sensemaking frameworks proposed by Klein et al. \cite{klein2007data} and Pirolli \& Card~\cite{pirolli2005sensemaking}. These frameworks incorporate two interconnected cyclic processes: (1) the \textit{Information Seeking/Elaboration} loop focuses on seeking out new information and filtering, aggregating, or transforming it into an accessible and understandable format; and (2) the \textit{Sensemaking/Reframing} loop involves analyzing and interpreting existing data, deciding whether additional data is required. If no further data is needed, it focuses on determining the most effective way to present the information

We designed GLOSS to directly address the barriers in sensemaking that exist in prior works (Table \ref{tab:system_limitations}). Thus, GLOSS works with high dimensional \textit{raw passive sensing data} from \textit{multiple sensor streams} and is capable of generating \textit{advanced code} to \textit{process and analyze} the data. It can be used \textit{off-the-shelf} (i.e., without any model training). GLOSS is a query-based system that can be used in human-facing tasks or automated background tasks. Moreover, GLOSS has \textit{minimum learning barriers} for human-facing applications as users can express their information needs as natural language queries. Additionally, GLOSS can \textit{personalize answers} based on an individual's needs.









\subsection{System Components and Flow}

\begin{figure}[h]
  \centering
  \includegraphics[width=0.90\linewidth]{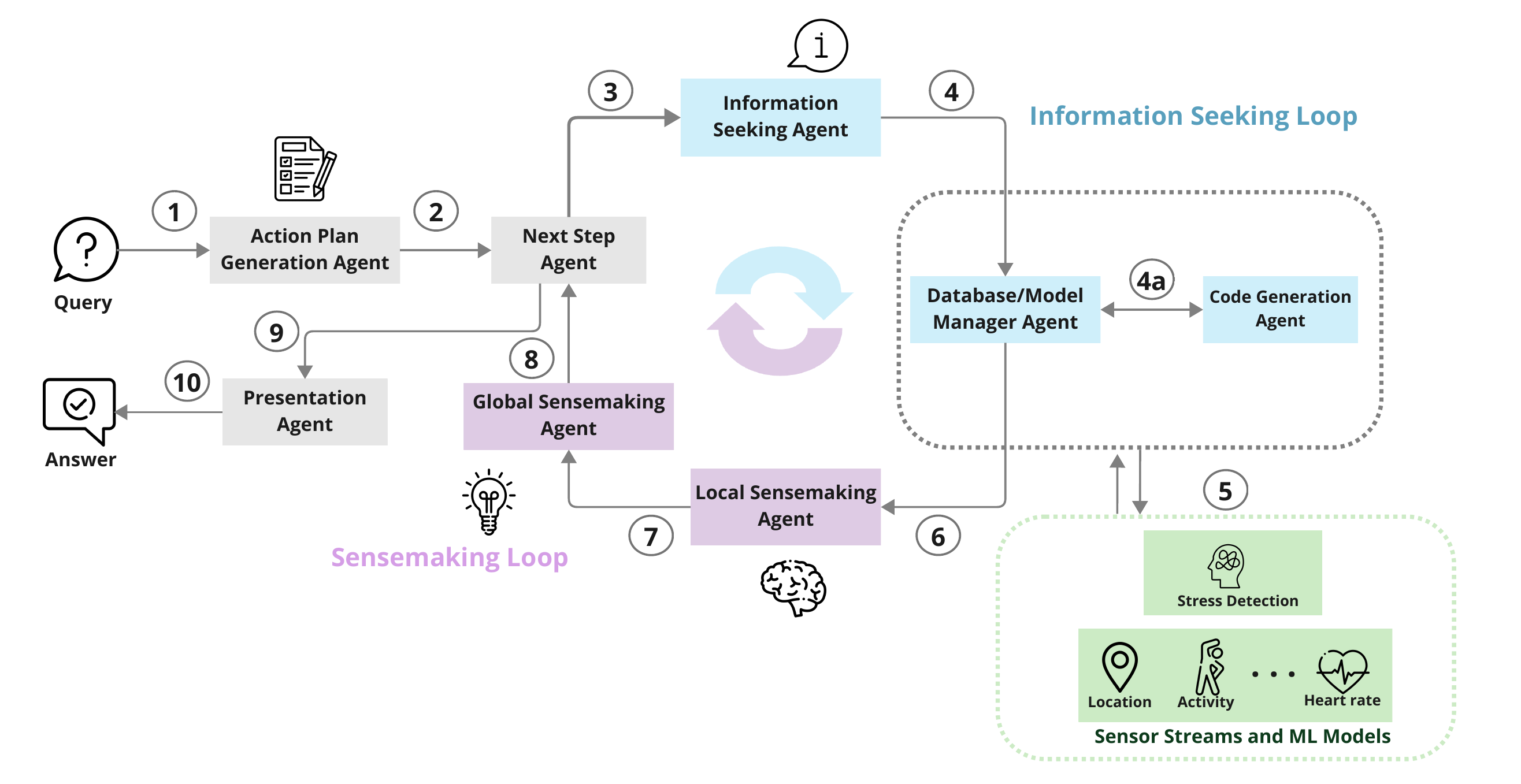}
  \caption{GLOSS system design where different LLM agents interact according to prior sensemaking theories}
  \label{fig:sys_des}
\end{figure}

\sethlcolor{user_query_hl}
We present the overall design of GLOSS in Figure~\ref{fig:sys_des}. To illustrate how GLOSS operates, we first define the necessary terminologies in Table~\ref{tab:sensemaking_term}.

\begin{table}[h]
\small
  \caption{Sensemaking Process Components}
    \renewcommand{\arraystretch}{1.5}
    \begin{tabular}{c|l|p{10cm}}
        \rowcolor{rowcolor1}
        \raisebox{-0.5\height}{\includegraphics[height=1.70em]{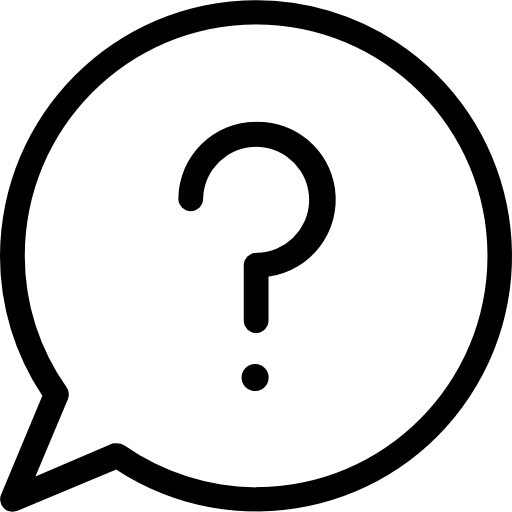}} 
        & \textbf{User Query} 
        & The natural language query that expresses the user's information need and initiates the sensemaking process. A user can be a human stakeholder or another automated system querying GLOSS.\\
        
        \rowcolor{rowcolor2}
       \raisebox{-0.5\height}{\includegraphics[height=2em]{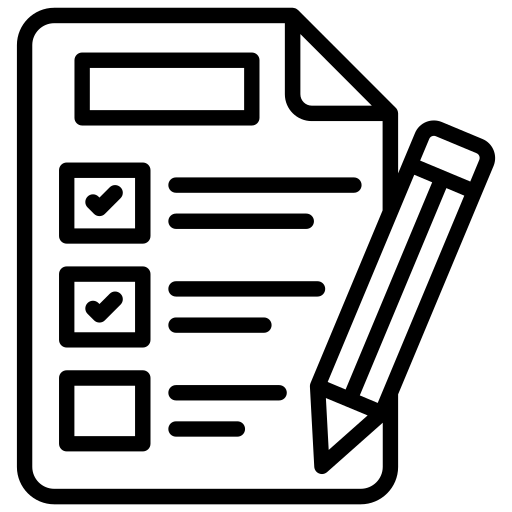}}
        & \textbf{Action Plan} 
        & A high-level strategy to drive the sensemaking process with the aim to address the \textit{user query}. \\
        
        \rowcolor{rowcolor1}
        \raisebox{-0.5\height}{\includegraphics[height=2em]{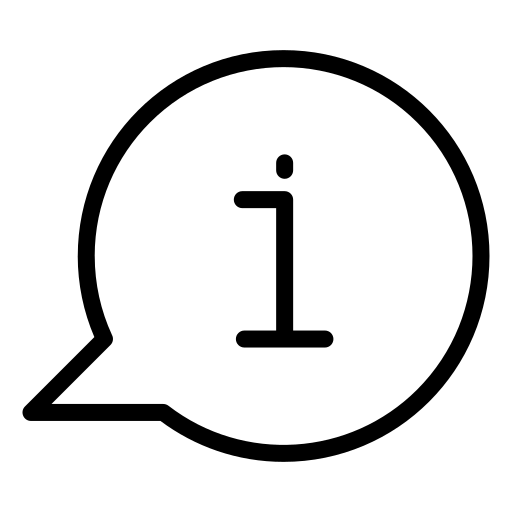}} 
        & \textbf{Information Request} 
        & A specific data retrieval request made to access databases or machine learning models available in GLOSS. \\
        
        \rowcolor{rowcolor2}
        \raisebox{-0.5\height}{\includegraphics[height=2em]{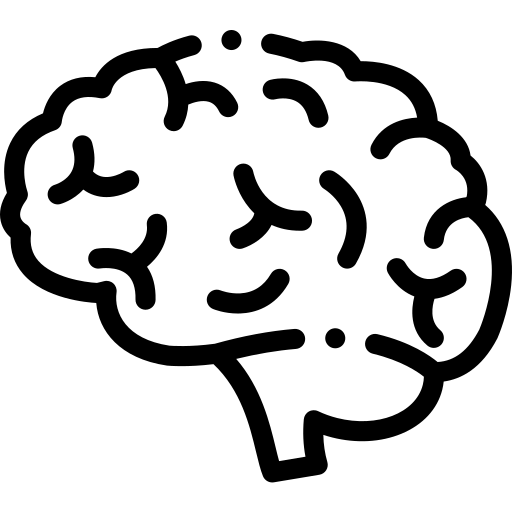}} 
        & \textbf{Memory} 
        & Memory contains all the data retrieved so far using \textit{information requests} to develop an \textit{understanding} of the user's query. \\
        
        \rowcolor{rowcolor1}
        \raisebox{-0.5\height}{\includegraphics[height=2em]{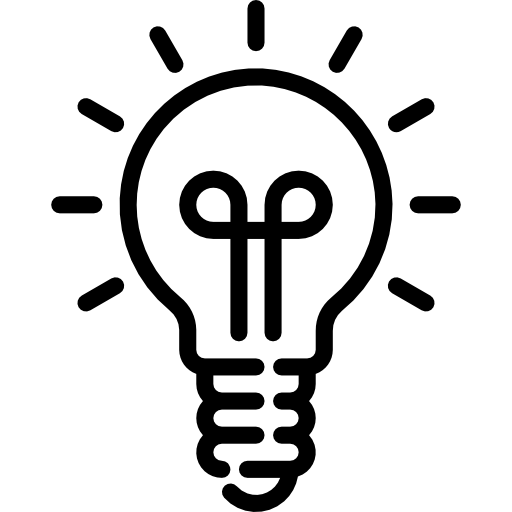}}
        & \textbf{Understanding} 
        & The interpretation or the answer to the \textit{user query} derived so far by the system based on data in the Memory. It also includes information on what additional data is needed to refine or expand this interpretation/answer. \\
        
        \rowcolor{rowcolor2}
        \raisebox{-0.5\height}{\includegraphics[height=2em]{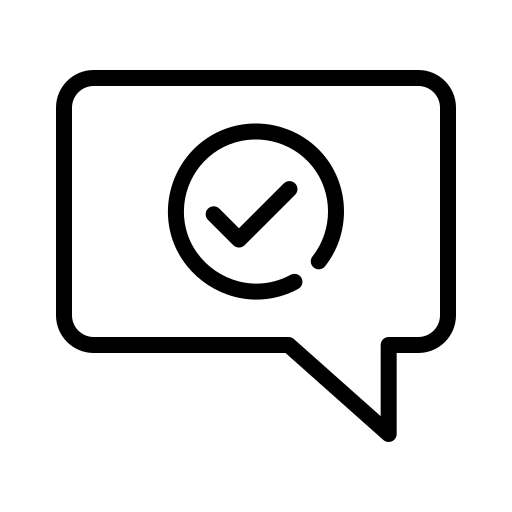}} 
        & \textbf{Answer} 
        & The answer to the query retrieved from \textit{understanding} and presented to the user according to their needs. \\
    \end{tabular}
    \label{tab:sensemaking_term}
\end{table}


In GLOSS, multiple LLM agents interact with each other with shared objective to interpret passive sensing data based on the \textit{user query}. The sensemaking process begins when the user submits the \textit{user query} expressing the information need (\circlednumber{1}). 

\subsubsection {Action Plan Generation}
Once GLOSS receives a \textit{user query}, the \textit{Action Plan Generation Agent} formulates a high-level \textit{action plan} to address the query using the system's access to available data and models. The \textit{action plan} provides a high-level, step-by-step guide to follow in order to derive an answer for the \textit{user query}. \rnr{In the prompt to the agent, we detail information} on available databases (sensor streams such as location data, Wi-Fi, Bluetooth, etc.) and models (e.g., a stress prediction model or an anomaly detection model). \rnr{The shared information includes a description of the contents of the database and the device used to collect the data (phone or smartwatch)}. The agent then uses this knowledge to devise an appropriate course of action. If the query cannot be answered with the available datasets, the agent halts the sensemaking process and informs the user; otherwise, it continues the process. \rnr{We show an example of an Action Plan when query requests data stream not currently supported by GLOSS in Figure \ref{fig:action_plan_and_understanding} (a)}.

\subsubsection{Information Seeking/Sensemaking Loop}

Following the action plan generation step, the sensemaking process transitions into the Information Seeking/Sensemaking Loop, beginning with an empty \textit{understanding} and \textit{memory} (\circlednumber{2}). At the start of this loop, the Next Step Agent uses \rnr{common sense reasoning capabilities of LLMs to evaluate} whether the current \textit{understanding} sufficiently addresses the \textit{user query} based on \textit{action plan} \rnr{generated in the previous step}. If it does, the process concludes; otherwise, it continues to the Information Seeking step.\rnr{The Next Step agent also halts the process if the understanding contains information that the data couldn't be fetched due to technical issues or data missingness.}
Since in the first iteration, \textit{understanding} is empty and \rnr{thus} cannot answer the \textit{user query}, the process moves to the information-seeking part of the loop (\circlednumber{3}). 

In this loop, the Information Seeking Agent \rnr{comes up with} specific \textit{information requests} with an aim to gather data relevant to developing a better \textit{understanding} of the \textit{user query}. \rnr{These requests can involve fetching and triangulating data from multiple databases}. The Information Seeking Agent makes these requests to Database/Model Manager Agent (DM Agent) based on the \textit{user query}, \textit{action plan}, \textit{memory}, and current \textit{understanding} (\circlednumber{4}). The DM Agent retrieves the relevant prebuilt data retrieval and processing helper functions based on the information requests. \rnr{For example, if an information request is "fetch me all GPS location values where activity is running," the DM agent retrieves all helper functions for location and activity databases.} \rnr{After retrieving helper functions}, DM Agent shares these functions along with their parameters, and their return values with the Code Generation Agent (\circlednumber{4a}). The Code Generation Agent creates Python or bash code to process the data as per the request made by the DM Agent. It uses helper functions to retrieve the data \rnr{from different databases relevant to information request} and generates additional code to \rnr{process and triangulate data to fulfill the request.} The agent makes multiple attempts to complete the request, attempting to resolve any programming errors on its own. The code generation \rnr{capabilities of LLMs} allows GLOSS to handle queries that involve basic to complex calculations, processing, \rnr{and data triangulations.}
The Code Generation agent then executes the code that can directly access raw data from databases or ML models through helper functions (\circlednumber{5}). After processing the data and computing results, the Code Generation Agent sends these results to the DM Manager Agent, who then forwards them to the sensemaking phase of the loop. 
(\circlednumber{6}).

In the sensemaking loop, the first step is local sensemaking. The results of the \rnr{code generated by the Code Generation Agent to complete the information request}  can be Python data structures (like lists, dictionaries, etc.), and the Local Sensemaking Agent's role is to generate a natural language representation of these results.
This step is crucial, as LLMs are more effective at processing data in natural language than in numerical formats or data structures \cite{lee_learning_2024, sui_table_2024}. Following this, the \textit{memory} is updated with the \textit{information request} and corresponding natural language representation of answer.
The process then advances to the global sensemaking step (\circlednumber{7}), where the Global Sensemaking Agent is responsible \rnr{for refining the previous \textit{understanding} of user query based on new information added to memory in the current iteration (in local sensemaking step)}. This agent reviews the \textit{action plan}, \textit{previous understanding}, and \rnr{all the data fetched so far} present in \textit{memory} to come up with a refined \textit{understanding} of the \textit{user query}. Moreover, while updating the \textit{understanding}, this agent \rnr{is instructed to include} information on any additional data that might be helpful \rnr{in completing/strengthening the \textit{understanding}} in the subsequent iterations. \rnr{Additionally, if in the current iteration, the data fetching and processing was unsuccessful due to technical issues or data missingness, the agent includes it in the \textit{understanding}}. \rnr{In Figure \ref{fig:action_plan_and_understanding} (b), we show an example \textit{understanding} when we forcefully restricted GLOSS's access to raw data to create a data retrieval failure.} After the global sensemaking step, the process goes back to the Next Step Agent again (\circlednumber{8}), completing one iteration of this loop. The loop continues until the Next Step Agent determines that the current \textit{understanding} sufficiently addresses the \textit{user query} \rnr{or there is information on some technical failure or data missingness in the \textit{understanding}.}

\subsubsection{Answer Presentation}

Once the Next Step Agent decides to conclude the process (\circlednumber{9}), the \textit{understanding} is handed off to the Presentation Agent. This agent 
\rnr {is instructed to} extract the \textit{answer} to \textit{user query} from the final \textit{understanding} and present it to the user according to specific presentation instructions or formats. This agent tailors the presentation to meet the specific needs of different users, as they may require the information in various formats. For example, human users can provide instructions to highlight qualitative aspects of well-being, or automated systems can request data in JSON outputs.

Overall, GLOSS mimics the iterative information-seeking/sensemaking process described by prior theories on human sensemaking \cite{klein2007data, pirolli2005sensemaking}. It retrieves necessary data and processes it, refining its understanding of the query in each iteration and finally presenting the answer to the user based on their needs.   

\begin{figure}[h]
\centering
\rnr{
\fbox{%
\begin{minipage}{0.38\linewidth}
\small

\textbf{Action Plan} \\
The query is regarding PPG (Photoplethysmogram) average value for \texttt{test006} on 2024-09-28. However, none of the provided databases contains PPG data. The available databases focus on activity, location, step count, heart rate, lock/unlock times, wifi connections, app usage, battery status, call logs, and stress levels. Since PPG data is not covered by any of these databases, the query cannot be answered with the given datasets.

\vspace{1em}
\hfill (a)
\end{minipage}}%
\hfill
\fbox{%
\begin{minipage}{0.57\linewidth}
\small
\textbf{Understanding} \\
To determine the total duration of each activity done by \texttt{test004} on 2024-07-12, we accessed the activity database for records associated with this user. Unfortunately, no activity data was found for \texttt{user\_id} 'test004' on the specified date of '2024-07-12'. As a result, it is not possible to calculate the total duration for each activity type such as stationary, automotive, cycling, walking, and running for that day. This absence of data could be due to the user not having any recorded activities on that date or an issue with data retrieval. If further data becomes available or if there is an alternative data retrieval method, it may be possible to revisit this query.

\vspace{1em}
\hfill (b)
\end{minipage}}%
}
\caption{\rnr{(a) Action plan when query requests data not supported by GLOSS (b) Understanding when GLOSS has technical issues.}}
\label{fig:action_plan_and_understanding}
\end{figure}

\subsection{Implementation Details}
We implemented GLOSS in Python using OpenAI's libraries, LangChain, and the AutoGen framework to create and manage interactions between various LLM agents. We used GPT-4o \cite{openai_gpt-4o_2024} as our Large Language Model and designed agent prompts following OpenAI's recommended best practices for prompt engineering \cite{openai_docs}. In our GPT-4o API calls, we maintained the default hyperparameter values ($temperature=1$ and $top_p=1$). To ensure security and prevent potential harm to the system running GLOSS, we executed any code generated by the LLM models within a Docker container.  Our GPT-4o deployment was HIPAA compliant, and OpenAI did not retain any queries for training their models. Moreover, we do not share large chunks of raw data with GPT as the code that GPT generates to access raw data is run in the local system running GLOSS, and then aggregated statistics are shared with the models. \rnr{To prevent a situation where information seeking/sensemaking process might run into an infinite loop we have we set up a hard cutoff of five iterations, after which GLOSS presents an answer to the user derived from the latest \textit{understanding}.  We have included sample runs of GLOSS to different queries in the supplementary materials.}

GLOSS supports multiple streams of passive sensing data while providing a scalable and modular framework for incorporating additional data streams. For our experiments and evaluations, we integrated several passive sensing data streams collected from iPhones and Garmin wearables. Details of the data streams are provided in Table \ref{tab:data_sources}. The raw passive sensing data was collected using an iOS application and stored in MongoDB databases. GLOSS can extract raw data directly from these MongoDB databases \rnr{through some pre-coded helper functions}. Our data collection protocol was approved by the Institutional Review Board (IRB) at our institution. \rnr {We can easily extend GLOSS to include additional data streams or ML models by just providing information about the data stream and some helper functions to fetch and retrieve data. We show sample databases and helper function descriptions we provide to agents in Figure \ref{fig:get_app_usage_blocks_and_databases}}.

\begin{table}[h!]
\centering
\renewcommand{\arraystretch}{1.5} 
\caption{Data sources, their descriptions, and frequency of collection.}
\begin{tabularx}{\textwidth}{l|X|r}
\hline
\textit{Data streams} & \textit{Description} & \textit{Sampling rate} \\ \hline \hline
Location & Provides GPS coordinates (latitude, longitude, and altitude) from the phone. & 1 minute \\ 
Activity & Provides activity categorizations such as stationary, walking, cycling, running, and automotive, using iOS's built-in activity detection. & Event-driven \\ 
App usage data & Provides open and close times of different iOS applications tracked through setting automations in the phone & Event-driven \\ 
Phone steps & Provides step counts, floors ascended/descended, and distance covered using phone sensors. & 1 minute \\ 
Phone lock/unlock & Tracks phone lock and unlock events. & Event-driven \\ 
WiFi & Provides connected/not connected status and WiFi details (WiFi name and SSID) when connected. & 1 minute \\ 
Call logs & Records incoming, outgoing, and missed calls. & Event-driven \\ 
Phone battery & Monitors changes in the phone's battery status. & Event-driven \\ 
Garmin steps & Tracks step counts from a Garmin smartwatch. & 1 minute \\ 
Garmin heart rate & Provides heart rate in bpm (beats per minute) from a Garmin watch. & 30 seconds \\ 
\makecell[tl]{Predicted Stress \\ Levels (ML model) } & Prediction of physiological stress (between 0:low stress and 1:high stress) using IBI and heart rate signals from Garmin watch based on the open-source model by Mishra et al.~\cite{mishra_evaluating_2020}. & 30 seconds \\ \hline
\end{tabularx}
\label{tab:data_sources}
\end{table}
\begin{figure}[h]
\centering
\fbox{%

\begin{minipage}{0.3\linewidth}
\rnr{
\small
\vspace{1em}
\textbf{App Usage Database:} \\
\textbf{Info:} Contains app usage data, including app names, open and close times, and durations. \\
\textbf{Device:} Phone
\vspace{1em}

\textbf{Garmin Stress ML Model:} \\
\textbf{Info:} Contains physiological stress predictions from ibi data recorded from the Garmin smartwatch. Physiological stress might not always be the same as psychological stress. The predictions are stress probabilities, with near 1 being more stressed. \\
\textbf{Device:} Garmin Smartwatch
\vspace{1em}}
\end{minipage}}%
\hfill
\fbox{%
\begin{minipage}{0.65\linewidth}
\small
\rnr{
\textbf{Name:} \texttt{get\_app\_usage\_blocks} 

\vspace{0.5em}
\textbf{Description:} 
Retrieves time blocks of app usage for a given user within a specified time range. Each block includes the app name, open time, close time, and duration.

\vspace{0.5em}
\textbf{Parameters:}
\begin{itemize}
  \item \texttt{uid} (string): The unique identifier for the user.
  \item \texttt{start\_time} (string): The start timestamp for the time range.
  \item \texttt{end\_time} (string): The end timestamp for the time range.
\end{itemize}

\textbf{Returns:} \\
A list of app usage blocks, each containing app name, open and close times, and duration in seconds.

\vspace{0.5em}
\textbf{Example:}
\begin{itemize}
  \item \{\texttt{'app'}: 'SnapChat', \texttt{'open'}: '2024-07-15 17:38:57', \texttt{'close'}: '2024-07-15 18:13:32', \texttt{'duration'}: 2075.0\}
  \item \{\texttt{'app'}: 'iMessage', \texttt{'open'}: '2024-07-15 19:07:34', \texttt{'close'}: '2024-07-15 19:08:12', \texttt{'duration'}: 38.0\}
\end{itemize}
}
\end{minipage}}%
\caption{\rnr{Example descriptions of databases (left) and helper functions (right) provided to agents for action plan and code generation.}}
\label{fig:get_app_usage_blocks_and_databases}
\end{figure}
\section{Method} 
\seclabel{meth}
GLOSS is an open-ended query-driven sensemaking system designed to explore and analyze passive sensing data.
In this section, we detail our evaluation method for the core GLOSS system using objective and subjective queries. Our aim here is to establish accuracy, consistency, and subjectivity in GLOSS's responses.












\subsection{Evaluation Queries}
 To evaluate GLOSS's performance, we crowd-sourced queries from seven researchers (one Postdoctoral scholar, five Ph.D. students, and one Masters' student) working in Ubiquitous Computing research and with prior experience with passive sensing data. We provided researchers with detailed information on the sensor data streams and machine learning models supported by GLOSS. We then instructed the researchers to formulate natural language queries to analyze the data. In the instructions, we asked researchers to draft open-ended queries for a hypothetical placeholder \textit{user ID} and \textit{date range}. Given the running cost and time of the system, we asked researchers to keep the date range up to seven days. We informed the researchers that their queries can ask for triangulating data across multiple datasets, such as, \textit{"How many outgoing calls were made when the user was connected to a wifi on 12/01/25?".} We include these instructions in our supplementary materials.

This approach resulted in a total of 122 human-generated natural language queries. For each human-generated query, we replaced the placeholder \textit{user ID} with two randomly chosen participants from our testing pool (12 participants, 913 participants-days). Additionally, the \texttt{date range} specified in the query was substituted with actual dates where we had data corresponding to the selected participants. For example, the query \textit{``Did userID have an extremely busy day on 26th October 24?''} was transformed into two distinct queries: \textit{``Did u010 have an extremely busy day on 11th June 23?'' }and\textit{ ``Did u013 have an extremely busy day on 25th May 23?''} By applying this approach, we doubled the original query set, resulting in a total of 244 queries, thus allowing us to effectively evaluate the accuracy and reliability of the system. We provide these queries in our supplementary materials.

Next, we followed the query categorizations defined by Merrill et al. ~\cite{merrill2024transforming} and began our coding process by adhering to their definitions of objective and subjective queries: 

\textbf{Objective queries:} Objective personal health queries are characterized by clear and specific answers. For example, the question, \textit{"On how many of the last seven days did the user exceed 5,000 steps?}" has a precise and tractable answer.

\textbf{Subjective queries:} Subjective queries are more open-ended and may have multiple valid answers. For example, \textit{“Was yesterday a productive day for the user?”} could yield different responses depending on how productivity is defined.

Two members of the team independently coded the queries, and any disagreements were resolved by consulting a third member of the team. During the coding process, we identified 14 queries that contained both objective and subjective components. Thus, we decided to go back and create a third category where we classified these queries as \textbf{mixed queries}. For instance,\textit{ "On Jan 5 2025, how many calls did user make while connected to Wi-Fi? Were they multitasking or catching up with someone important?"} has both objective and subjective components. Out of a total of 244 queries, we classified 80.32\% (196) as objective, 13.93\% (34) as subjective, and 5.73\% (14) as mixed queries. 

\subsection{Evaluation Metrics}
Our goal was to compare GLOSS with similar approaches in prior works, but those approaches are close-sourced. Moreover, several of those approaches do not have the capabilities to deal with completely open-ended queries or multi-modal data triangulation. Thus, to compare and evaluate GLOSS, we implemented a baseline model based on a widely used Retrieval-Augmented Generation (RAG) technique in prior works in passive sensing \cite{yang_drhouse_2024, yang_retrieval-augmented_2025}. Similar to prior works, We transformed raw sensor data from different sensor streams into a natural language format \cite{li2024vital, kim_health-llm_2024, llmsense}. This natural language data is stored in a Chroma database and retrieved based on the user query. We used GPT-4o for the implementation and used LangChain's RAG framework \footnote{https://python.langchain.com/docs/concepts/rag/} to implement our baseline RAG model. 

Next, we designed separate evaluation metrics for objective and subjective queries. For mixed queries, we evaluated the objective components using objective metrics and the subjective components using subjective metrics.
For objective queries, we ran queries in both GLOSS and RAG, setting presentation instructions in GLOSS to \textit{``answer clearly and concisely''}.
we evaluated both GLOSS and the RAG in terms of \textit{accuracy} and \textit{consistency}. Accuracy is measured by evaluating whether the model retrieved the correct answer based on the underlying logic it followed. Consistency was measured by passing the same query to the model three times; a response was deemed consistent if it remained identical across all three runs. Some prior works used template-based queries that only covered certain operations and calculations on aggregated data, making automatic evaluation of accuracy possible. On the other hand, our queries were completely open-ended, requiring the direct use of raw data and the triangulation of data. Thus, to evaluate the accuracy of models, we needed to manually write Python code and analyze the data based on the logic model followed. Our queries spanned over some simple tasks like \textit{``names of all Wi-Fi networks u010 connected on 11th June 23 between 10 am and 7 pm?''} to some complex queries like \textit{``did u010 do any outdoor exercising between 2023-06-11 and 2023-06-17?''} requiring triangulating location, activity, and step data. Thus, we randomly sampled 60 queries \footnote{The query dataset size of some prior works \cite{merrill2024transforming} is large compared to ours as their objective queries are designed using templates on aggregated data and thus afford automatic evaluation. In our case, the objective queries were open-ended and required domain experts to manually write code for each query to determine accuracy.} from a total of 210 queries (196 objective queries + 14 mixed queries), and the first author wrote code to process \& analyze the date to calculate accuracy for these 60 queries. Consistency, however, was evaluated across all 210 queries. 

For subjective queries, we generated responses from both GLOSS and RAG, setting presentation instructions in GLOSS to \textit{``explain clearly and in detail''}. We presented the responses to two annotators as \textit{response 1} and \textit{response 2}, ensuring that the source of each response remained hidden. To avoid bias, both \textit{response 1} and \textit{response 2} included a random number of GLOSS and RAG-generated responses. Following \cite{merrill2024transforming}, the annotators were tasked to evaluate each model response based on the following attributes: Relevance (relevance of the data used), Interpretation (accuracy in interpreting the question), Domain Knowledge (application of domain knowledge), Logic (logical correctness), and Clarity (clarity of communication). Additionally, they rated the overall reasoning of each response on a Likert scale from 1 (``Poor'') to 5 (``Excellent''). For the subjective query evaluations, we set the presentation instructions in GLOSS to \textit{``explain clearly and in detail''}.

\section{Results} 
\seclabel{related}
In this section, we report the performance of GLOSS on both objective and subjective queries, comparing it with a RAG-based baseline (RAG)

\begin{wraptable}{r}{0.4\textwidth}
\small
\centering
\renewcommand{\arraystretch}{1.2} 
\caption{Comparison of model accuracy and consistency of GLOSS and RAG on objective queries}
\begin{tabular}{|p{0.08\textwidth}|p{0.1\textwidth}|p{0.1\textwidth}|}
\hline
\textit{Model} & \textit{Accuracy  \%(n)} & \textit{Consistency \%(n)} \\ \hline \hline
GLOSS & 87.93\% (60) & 66.19\% (210)\\ 
RAG & 29.31\% (60)& 52.85\%  (210)\\ \hline
\end{tabular}
\label{tab:model_comparison}
\end{wraptable}

\subsection{Objective Evaluation}
In Table \ref{tab:model_comparison}, we show the accuracy and consistency comparison between GLOSS and RAG. We performed paired t-tests to measure the significance of accuracy and consistency performances. GLOSS significantly outperforms RAG in terms of accuracy ($t(59) = 8.24, p < .001$). We observe this notable difference due to GLOSS's ability to generate code for calculations and data processing, whereas the RAG-based approach relies solely on LLM calls to interpret the data. This aligns with previous studies indicating that LLMs and RAG-based approaches struggle with mathematical tasks and calculations \cite{agrawal_mindful-rag_2024, lewkowycz_solving_2022}.
The consistency of GLOSS  (66.19\%) is significantly higher when compared to RAG (52.85\%) ($t(59) = 3.33, p = .001$). GLOSS’s consistency, however,  is lower than its accuracy, primarily due to ambiguity in the queries. Some queries labeled as objective queries by annotators had a single tractable answer provided a particular logic but had multiple valid logical paths. While GLOSS accurately computed the answer, it followed different logical paths in different runs. For instance, for query, \textit{"What day of the week on the first week of June that u010 has the highest mobility?"}, GLOSS used a high step count as an indicator of mobility in one run, while in another, it relied on the maximum difference between traveled GPS coordinates.

\begin{wrapfigure}{r}{0.42\textwidth}
  \vspace{-50pt} 
  \centering
  \includegraphics[width=0.90\linewidth]{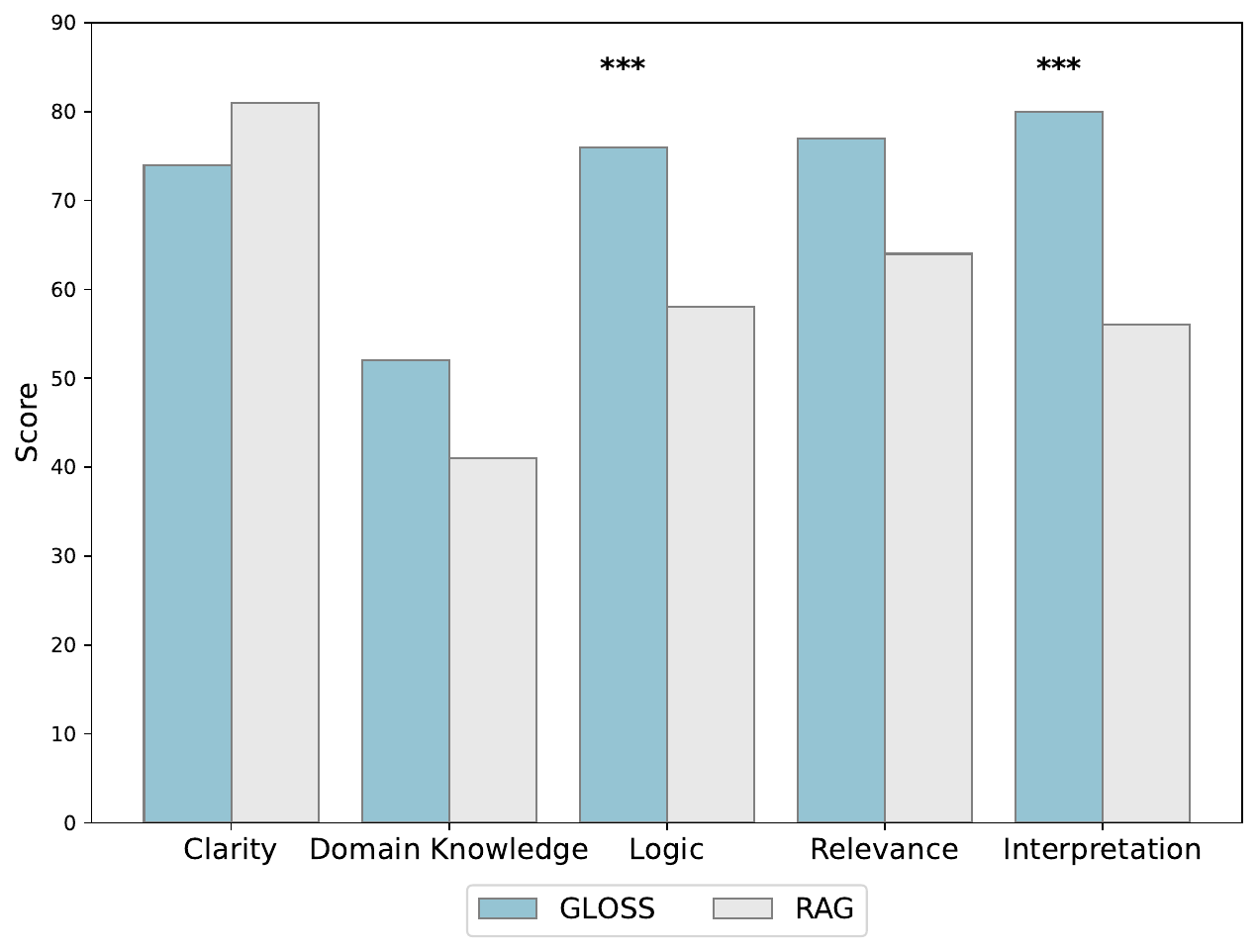}
  \caption{Comparison of GLOSS and RAG across multiple dimensions. We found statistical evidence between the subjective measurements of GLOSS and RAG in terms of logic and interpretation (highlighted with *** on top of the bars)}
  \label{fig:subj_comp}
  \vspace{-10pt} 
\end{wrapfigure}

\subsection{Subjective Evaluation}
The trends in subjective evaluations were more subtle compared to objective evaluations. In this subsection, we present our results across five different criteria. We summarize these trends in Figure \ref{fig:subj_comp}.

\subsubsection{Clarity} 
To rate \textit{Clarity}, we asked the annotators: `\textit{`Is this response easy to read and is it clear?''}. Overall, the annotators found that RAG was able to provide more concise, readable responses compared to GLOSS. However, we found no statistically significance between GLOSS and RAG responses ($\chi^2(1)=1.59, p = .21$). We show the comparison between RAG and GLOSS responses in terms of clarity in  Table \ref{tab:clarity}.  In the example given in Table \ref{tab:clarity}, GLOSS printed out all timestamps with a heart rate greater than 120 bpm, making the response hard to read and less actionable. It is, however, important to note that RAG demonstrated significantly lower objective accuracy (Table~\ref{tab:model_comparison}), and often led to hallucinated and made-up numbers and inferences that did not actually exist in the data.
\begin{table}[h]
\small
\renewcommand{\arraystretch}{1.2}
\caption{Comparison between RAG and GLOSS responses in terms of \textbf{clarity}. Both models understood the question and were able to come up with a relevant response. Due to its data-driven approach, GLOSS tends to present data in a longer format, while RAG tends to provide a shorter response with \emph{incorrect/made-up data}.}
\begin{tabular}{p{0.1\textwidth} p{0.85\textwidth}}
\hline
\multicolumn{2}{l}{\textit{Comparison between GLOSS and RAG}} \\
\hline \hline
\textbf{Query} & On June 4, 2023, which Wi-Fi network was u013 connected to when their heart rate went over 120 bpm? Could it have been during an exciting or intense moment? \\  
\hline
\textbf{RAG} & On June 4, 2023, u013's heart rate went over 120 bpm at multiple instances: \\
& 1. From 12:33:37 onwards, their heart rate reached a peak of 134 bpm, and during this time their phone was connected to a Wi-Fi network labeled \textcolor{maroon}{"not connected." - \textit{made-up network}} \\  
& 2. Starting from 13:31:39, their heart rate again exceeded 120 bpm and continued to increase. By 13:33:08, their phone was connected to the Wi-Fi network "[anonymized network]." Shortly after at 13:45:39, their heart rate was recorded at lower levels. \\  
& Given the spikes in heart rate and the records around those time slots, it could suggest that the increases in heart rate were due to some exciting or intense moment, especially looking at the second instance on the "[anonymized network]." network. \\  
& \textbf{Summary:} \\
& - u013's heart rate exceeded 120 bpm when connected to the "not connected" network, between 12:44:37 and 12:54:07. \\  
& - Another spike was recorded while connected to the "[anonymized network]" network around 13:31:39. \\  
& Without further data, we can reasonably infer that the elevations in heart rate could be related to some form of physical or emotional stress. \\  
\hline
\textbf{GLOSS} & On 2023-06-04, user u013 experienced several instances where their heart rate exceeded 120 bpm at the following times: \\  
& - 10:52:35 \\  
& - 10:53:05 \\  
& - 10:53:35 \\   
& ... \textcolor{blue}{(redacted 92 lines) --- \textit{overly verbose response}} \\  
& During these periods of elevated heart rate, user u013 was mostly connected to the "[anonymized network]." Wi-Fi network. However, from 12:38:07 to 13:13:08, user u013 was not connected to any Wi-Fi network. Due to an inability to fetch stress level data, it is not possible to definitively determine whether these heart rate spikes were due to exciting or intense moments. \\ \hline 
\end{tabular}
\label{tab:clarity}
\end{table}
\begin{table}[h]
\small
\renewcommand{\arraystretch}{1.2}
\centering
\caption{Comparison between RAG and GLOSS responses in terms of \textbf{relevance}. Both models understand the question and was able to come up with a response. GLOSS, however, searched for evidence using multiple sensor streams before making the final answer, while RAG answered using only two sensor streams - location and step count.}
\begin{tabular}{p{0.2\textwidth}|p{0.75\textwidth}}
\hline
\multicolumn{2}{l}{\textbf{Comparison between GLOSS and RAG}} \\ \hline
\textbf{Query} & Will test007 be more likely to step out of the house on 10/16/24 based on past two days of data? \\ \hline
\textbf{RAG} & Based on the provided sensor data for the past two days, it appears that test007 \textcolor{maroon}{has been stationary and has not recorded any steps or movement.} \\  
& On both October 14th and October 15th, the data shows that test007 remained \textcolor{maroon}{in the same locations} without any significant activity. \\  
& Given this pattern of inactivity, it is unlikely that test007 will step out of the house on October 16, 2024. \\ \hline
\textbf{GLOSS} & Based on the data collected for user\_id test007 for the dates 2024-10-14 and 2024-10-15, we observe the following insights: \\  
& \textbf{Activity Data}: The user was primarily stationary with periodic walking intervals. For example, on 2024-10-14, there was notable stationary time from 12:21 AM to 9:36 AM with brief walking episodes. \\  
& \textbf{Location Metrics}: The user visited 10 different significant locations with substantial displacements. The maximum displacement recorded was 18,758.52 meters, and the total travel distance was 36,829.64 meters over the two days. \\  
& \textbf{Steps Data from Phone}: The user took a total of 3,844 steps, covering 2,783.15 meters, and ascended 10 floors while descending 9 floors. \\  
& \textbf{Wi-Fi Connection Logs}: The user connected and disconnected from multiple Wi-Fi networks, such as [anonymized networks], indicating movement and not being in a single location for extended periods. \\  
& \textbf{Phone Lock/Unlock Times}: The phone was locked 123 times and unlocked 124 times, with the phone being unlocked for a total of approximately 19 hours and 21 minutes. \\  
& This comprehensive data, showing consistent \textcolor{blue}{activity levels, movement patterns, Wi-Fi connections, and phone usage}, suggests that the user had a pattern of stepping out over the past two days. Therefore, it is likely that user test007 will step out of the house on 2024-10-16. \textcolor{blue}{--- \textit{triangulating multiple database to draw conclusions}} \\ \hline
\end{tabular}
\label{tab:relevance}
\end{table}

\subsubsection{Relevance}
 Relevance was rated as: \textit{``Does this response reference all relevant user data?''}. Table \ref{tab:relevance} shows the comparison between RAG and GLOSS responses in terms of relevance. Overall, GLOSS is slightly better at triangulating multiple data sources to answer the query compared to RAG. We found, however, that there is no statistically significance between GLOSS and RAG responses in terms of relevance ($\chi^2(1)= 3.44, p = .06$) In the example shown in Table \ref{tab:relevance}, GLOSS investigated various sensor streams, including step count, location, Wifi, and phone usage pattern, before coming up with the conclusion. In contrast, RAG came to the conclusion after only looking at location and step count data.

\begin{table}[h]
\small
\renewcommand{\arraystretch}{1.2}
\centering
\caption{Comparison between RAG and GLOSS responses in terms of \textbf{interpretation}.}
\begin{tabular}{p{0.1\textwidth}p{0.85\textwidth}}
\hline
\multicolumn{2}{l}{\textit{Comparison between GLOSS and RAG}} \\ \hline\hline
\textbf{Query} & On Aug 1 2024, how many steps did test006 take while connected to wi-fi? it might show moments of movement at home or work. \\ \hline
\textbf{RAG} & On August 1, 2024, while connected to Wi-Fi, test006 took the following steps: \\  
& \textbf{From 08:42:46 to 10:37:53 (connected to <institution wifi name>):} \\  
& - 83 steps between 08:49:59 and 08:51:06. \\  
& \textbf{From 20:48:02 to 21:01:02 (connected to not connected):} \textcolor{maroon}{wrong inferencing thinks ``not connected'' is a wifi name}\\  
& - 298 steps between 20:49:11 and 20:52:17. \\  
& - 38 steps between 20:53:13 and 21:04:45 (partially overlapping with the next Wi-Fi connection window). \\  
& Adding these up: \\
& 83 steps (<institution wifi name>) + 298 steps (not connected) + 38 steps (partially in the not connected window) = 419 steps. \\  
& Therefore, test006 took \textbf{419} steps while connected to Wi-Fi on August 1, 2024. \\ \hline
\textbf{GLOSS} & On August 1, 2024, test006 took a total of \textbf{9325} steps while connected to Wi-Fi. Here are the details: \\  
& \textbf{Connected to <institution wifi name>:} \\  
& - From 05:05:19 to 08:31:46, took 1079 steps \\  
& - From 08:42:46 to 10:37:53, took 1280 steps \\  
& \textcolor{blue}{<redacted>} \\  
& \textbf{Connected to 'xfinitywifi':} \\  
& - From 16:00:25 to 17:03:58, took 581 steps \\  
& \textcolor{blue}{<redacted>}
\\
& In total, test006 took \textbf{9325} steps while connected to Wi-Fi on 2024-08-01. \\ \hline
\end{tabular}
\label{tab:interpretation}
\end{table}

\subsubsection{Interpretation}
. Interpretation is rated as \textit{``Does this response properly interpret referenced user data?''}. Table \ref{tab:interpretation} shows an example of the comparison between RAG and GLOSS in terms of interpretation. GLOSS was statistically better at interpreting the data before arriving at a conclusion compared to RAG ($\chi^2(1)= 12.69, p<.001$).

In the example shown in Table \ref{tab:interpretation}, both RAG and GLOSS understood the question and provided evidence for the responses based on data. RAG, however, made a false interpretation of the data at the end of the response. RAG inferred ``not connected'' as a wifi network, whereas it meant the phone was not connected to any wifi network at that time.

\begin{table}[h]
\small
\renewcommand{\arraystretch}{1.2}
\centering
\caption{Comparison between RAG and GLOSS responses in terms of \textbf{domain knowledge}. In this example, RAG was unable to notice the abnormality in Garmin step count data, and arrived at the wrong conclusion. GLOSS, however, was able to explain the discrepancy between the phone and Garmin step count using common knowledge.}
\begin{tabular}{p{0.1\textwidth}p{0.85\textwidth}}
\hline
\multicolumn{2}{l}{\textit{Comparison between GLOSS and RAG}} \\ \hline \hline
\textbf{Query} & Did test008 a physically active person in general based on data from 06/25/24 to 06/29/24? \\ \hline
\textbf{RAG} & Based on the data provided from 06/25/24 to 06/29/24, test008 does not appear to be a physically active person during this period. \\  
& The data shows that test008 consistently walked 0.0 steps, covered 0.0 meters, and remained stationary for multiple periods each day. \\  
& There is no record of physical activity in the form of steps taken or floors climbed or descended. Therefore, it can be concluded that test008 was not physically active during the specified dates. \\ \hline
\textbf{GLOSS} & Based on the data from June 25, 2024, to June 29, 2024, user test008 appears to be generally physically active. \\  
& \textbf{Activity Data:} \\  
& - Stationary: 101 times for a total of 4 days, 12 hours, and 49 minutes \\  
& - Walking: 72 times for a total of 5 hours and 42 minutes \\  
& - Automotive: 65 times for a total of 1 hour and 52 minutes \\  
& - Running: 10 times for a total of 15 minutes \\  
& - Cycling: 59 times for a total of 2 hours and 37 minutes \\
& \textbf{Phone Steps Data}: \\  
& - Total Steps: 32,688 \\  
& - Average Steps per Day: 6,537.6 \\
& \textbf{Garmin Steps Data}: \\  
& - Total Steps: 0 \\  
& - Average Steps per Day: 0 \\  
& Despite the zero steps recorded on the Garmin device \textcolor{blue}{(likely due to not wearing the device or data retrieval issues)}, the variety and duration of physical activities logged and phone steps data indicate that test008 was engaging in physical activities and was generally physically active during this timeframe. \textcolor{blue}{\textit{ --- ability to give explanation for abnormal data}} \\ \hline
\end{tabular}
\label{tab:domain_knowledge}
\end{table}

\subsubsection{Domain Knowlegde}
Table \ref{tab:domain_knowledge} shows an example of the comparison between RAG and GLOSS in terms of domain knowledge. Interpretation is rated as \textit{``Does this response contain evidence of relevant
and correct domain knowledge (e.g., mention of
a relevant and/or correct fact for answering the
question)?''}. Overall, GLOSS was slightly better at interpreting the data before arriving to conclusion compared to RAG, especially in query that requires the use of common sense knowledge. We found, however, that there is no significance between GLOSS and RAG responses in terms of domain knowledge ($\chi^2(1)= 1.9, p = .17$).

In the example shown in Table \ref{tab:domain_knowledge}, both models understood the question and referenced relevant data to answer. However, RAG failed to detect the abnormality in the Garmin step count data and reached the wrong conclusion. In contrast, GLOSS not only identified the discrepancy between the phone and Garmin step count but also explained it using domain knowledge. This highlighted GLOSS' ability to incorporate data from multiple sensors and arrive at the correct conclusion.

\begin{table}[h]
\small
\renewcommand{\arraystretch}{1.2}
\centering
\caption{Comparison between RAG and GLOSS responses in terms of \textbf{logic}. In this example, RAG was making a minor calculation mistake at the end of its response.}
\begin{tabular}{p{0.1\textwidth}p{0.85\textwidth}}
\hline
\multicolumn{2}{l}{\textit{Comparison between GLOSS and RAG}} \\ \hline
\textbf{Query} & Does test006 prefer texting over calling based on data from 06/08/24 to 06/13/24? \\ \hline
\textbf{RAG} & Based on the data provided from 06/08/24 to 06/13/24: \\  
& \textbf{Phone call durations:} \\  
& - 06/08/24: 3 seconds, 29120 seconds (approximately 8.09 hours) \\  
& - 06/09/24: 5 seconds, 532 seconds (approximately 8.87 minutes), 2 seconds, 45 seconds, 3 seconds \\  
& - 06/11/24: 2 seconds, 293 seconds (approximately 4.88 minutes) \\  
& \textbf{iMessage durations:} \\  
& - 06/08/24: 3 seconds, 0 seconds, 0 seconds, 9 seconds \\  
& - 06/09/24: 5 seconds, 19 seconds, 9 seconds, 0 seconds, 60 seconds, 0 seconds, 9 seconds, 10 seconds \\  
& - 06/11/24: 102 seconds (approximately 1.7 minutes), 19 seconds, 113 seconds (approximately 1.88 minutes), 7 seconds \\ 
& \textbf{Summarizing the total duration spent on phone calls versus iMessage:} \\  
& - Total phone call time: \textcolor{maroon}{29120 + 5 + 532 + 2 + 45 + 3 + 2 + 293 = 29902 seconds (approximately 8.3 hours) --- \textit{wrong calculation, should be 30,002 seconds. The overall logic remains wrong even if this calculation was right.}} \\  
& - Total iMessage time: 3 + 0 + 0 + 9 + 5 + 19 + 9 + 0 + 60 + 0 + 9 + 10 + 102 + 19 + 113 + 7 = 365 seconds (approximately 6.08 minutes) \\  
& Considering the significant difference in time spent, it is evident that test006 prefers calling over texting. The total duration spent on phone calls far exceeds the time spent using iMessage. \\ \hline
\textbf{GLOSS} & Based on the data available, for user\_id 'test006' between 2024-06-08 and 2024-06-13, the total duration of usage for messaging apps is 12 hours, 51 minutes, and 33 seconds, while the total duration of calls made and received is 37 hours, 37 minutes, and 6 seconds. \\  
& This indicates that during this period, test006 spent significantly more time on phone calls than on texting. Therefore, test006 prefers calling over texting during the specified date range. \\ \hline
\end{tabular}
\label{tab:logic}
\end{table}

\subsubsection{Logic}
Table \ref{tab:logic} shows an example of the comparison between RAG and GLOSS in terms of logic. Logic is rated as \textit{``Does this response contain evidence of correct reasoning
and logic?''}. Overall, GLOSS was statistically better at reasoning and logical conclusion compared to RAG, especially in query that requires calculation of quantitative data ($\chi^2(1)= 6.64, p = .001$).

In the example shown in Table \ref{tab:logic}, both models understood the question, referencing relevant data, and interpreting the data correctly. RAG, however, made a calculation mistake in its response, possibly due to hallucination. GLOSS, on the other hand, never uses LLM for calculation but relies on code generation, which reduces the likelihood of making calculations and logical mistakes.

\section{Potential Use Cases of GLOSS}
\label{sec:usecase}

GLOSS provides a powerful system for analyzing and understanding passive sensing data, transforming what is often considered a complex and time-consuming process into a more accessible and efficient one. Its open-ended sensemaking capabilities make GLOSS a promising solution for various applications in personal health informatics (PHI) and well-being monitoring. In this section, we present multiple potential use cases of GLOSS, drawing inspiration from prior and ongoing work in UbiComp and the HCI communities. While the scope of this paper focuses on the core GLOSS system with demonstrations of potential use cases, we plan to conduct a thorough evaluation of each of these use cases as part of our future work. 

\begin{figure}[h]
  \centering
  \includegraphics[width=0.90\linewidth]{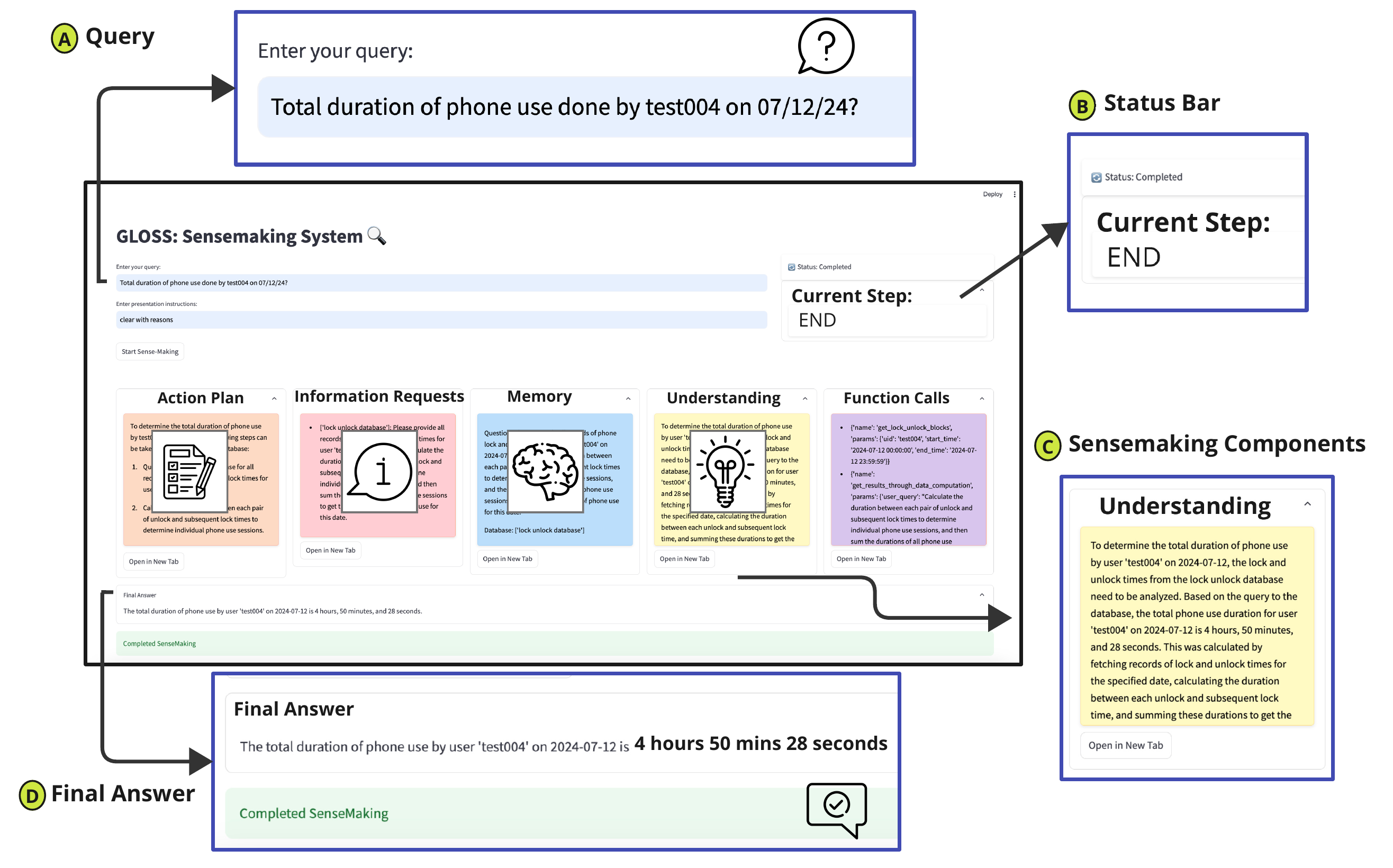}
  \caption{\rnr{GLOSS UI design that shows relevant components of sensemaking (such as Action Plan, Information Requests, etc.). As the sensemaking process continues, these different components in the UI get updated with relevant information.}}
  \label{fig:glosui}
\end{figure}

To make a user-friendly way to interact with GLOSS, we implemented \textit{a chat-based user interface (UI)} for GLOSS using the Streamlit Python library. The GLOSS UI displays different components of sensemaking like \textit{action plan}, \textit{information requests}, \textit{memory}, and \textit{understanding} \rnr{updating them} as the iterative sensemaking process continues. We present the GLOSS UI in Figure \ref{fig:glosui}. \rnr{The user begins by entering a query and presentation instructions into the search box at the top of the screen (\circlelabel{A}). The sensemaking process starts when the user presses ``Start Sense-Making'' button. Throughout the process, the status bar (\circlelabel{B}) shows the current state of the sensemaking process (e.g., information seeking, local sensemaking, code generation, etc.) As the process continues, different components of sensemaking (\textit{action plan}, \textit{information requests}, \textit{memory}, and \textit{understanding}) are updated (\circlelabel{C}). Finally, when the sensemaking process finishes, the final answer is presented to the user tailored according to their presentation instructions  (\circlelabel{D}).}

\subsection{Supporting Non-Computer Science researchers with Sensemaking}

With the ubiquitous presence of mobile phones and wearables in people’s lives, behavioral scientists, clinicians, and psychologists have shown a growing interest in passive sensing data to study people's behavioral patterns \cite{anikwe2022mobile, place2017behavioral}. \rnr{Passive sensing has the potential to unobtrusively collect and reveal meaningful affective and behavioral patterns over time ~\cite{abdullah2018sensing, mohr2017personal}}. These researchers, \rnr{although interested in passive sensing data}, often lack computer science training and experience (we call them non-CS researchers). \rnr{This lack of expertise, creates} substantial barriers for \rnr{non-CS researchers} in independently processing, analyzing, and interpreting longitudinal passive sensing data \cite{felix2019mobile}. As a result, these researchers often depend on the expertise of trained computer scientists and data analysts to process and interpret passive sensing data \cite{xu2021understanding}. While this collaboration is valuable, it can also lead to challenges. For instance, if non-CS researchers require additional data or specific analyses, they must request assistance from their computer science collaborators and/or staff. This process can often be time-consuming, leading to considerable delays and increased effort for everyone.

GLOSS provides a promising solution to this barrier faced by non-CS researchers. Its chat interface empowers them to interpret and analyze passive sensing data independently. GLOSS  can generate code, perform data triangulation, and present insights in a transparent and interpretable format customized to their specific needs. 
To provide a concrete example of GLOSS’s potential impact, we conducted semi-structured interviews and a think-aloud session with three psychologists from a research group working on passive sensing data to understand adolescents' mental health and well-being at a private R1 institution. We set up GLOSS with a subset of their study data so they were acutely familiar with their dataset and could have an effective interaction with GLOSS. Each interview session lasted for 30-45 mins. We started the interview by asking questions about their current practices in dealing with passive sensing data, followed by a tutorial demonstrating how to use GLOSS through its user interface. Next, in a think-aloud session, participants interacted with GLOSS, which lasted for around 10-15 minutes. We asked participants to come up with open-ended queries based on data monitoring and analysis tasks they encounter on a day-to-day basis. After the session, we asked participants questions about their experience and perceived usefulness of GLOSS. At the end of the interview, participants completed a modified version of the System Usability Scale (SUS). This study was approved by our IRB, and participants were compensated \$15 for their time. 

We conducted the meetings in person and recorded the audio and screen interaction using Zoom. The audio from the meetings was automatically transcribed using Microsoft Word transcriber. Two members of our research team independently coded the interviews using the Open Coding approach ~\cite{corbin1990grounded}. The coders later met to coalesce common themes and discuss disagreements. We present the main themes that emerged from the interview sessions below:

\subsubsection{Current analysis and monitoring methods:} All the participants relied on their existing study dashboards for data visualization and monitoring. P3 explained, \textit{``Right now our team has a dashboard that allows us to see different aspects of different data streams we're collecting.''}  Our participants did not have formal training in computer science (mean self-reported coding proficiency was 3.34/10). Some participants mentioned that they use R for statistics and calculation sometimes but would not consider them proficient. Due to a lack of technical expertise, participants mentioned that accessing data not available on the dashboard requires reaching out to technical team members. They described this process as often time-consuming and requiring considerable effort. As P2 said, \textit{``I would have to reach out to the data scientists and computer scientists that our team works with and ask them for support (for getting additional data) because I'm not necessarily sure what the best way to go about that would be''}. Even after receiving data from technical staff, participants mentioned that they often need to manually review and, at times, hand-code data, which is a highly labor-intensive process. P1 elaborated on this, sharing, \textit{“Yes (it is time-consuming), especially when I'm hand coding when I know it would be a lot easier not to hand code, but it feels easier than having to learn how to navigate some CS system. It is scary.''}. Participants expressed that this manual coding method would not be scalable if they had a large participant pool in their studies. 

\subsubsection{Experience with GLOSS:} 

Participants found GLOSS to be a valuable tool for monitoring and analyzing data, highlighting its ability to save significant amounts of time. P2 shared, ``\textit{Overall, it was a really amazing experience. I can already see it saving so much time. Maybe a minute from typing this to seeing the result [...], and it’s absolutely miraculous that we can do it this quickly.}'' They also felt that GLOSS made complex tasks much more manageable. GLOSS was described as easy to use, with P1 noting, ``\textit{Since you’re able to just kind of ask the question however it pops up in your head, you don’t really have to think through the exact code or anything to figure it out. It makes it much more accessible.}'' Similarly, P2 remarked, “\textit{You can use it (GLOSS) like a Google search. People are very comfortable with that.}''

Participants also appreciated GLOSS's user interface, which displayed various components such as the \textit{action plan}, \textit{information requests}, \textit{memory}, and \textit{understanding}. They felt that showcasing how GLOSS performs sensemaking through these elements improved the system's explainability and strengthened their trust in it. P2 expressed, \textit{``It's nice that it's (GLOSS) just so transparent because I feel like that's a big issue that people talk about with these things, with language learning models. You'll get this output, but then it's like, OK, how did we get here? Can we really trust it? For things like research, when you need to be so precise and accurate, it's really helpful that we can see each step it went through.'}' The SUS survey reaffirmed that participants viewed GLOSS as a valuable, easy-to-use, and easy-to-learn tool, with an average aggregated SUS score of 84.02\%

\subsubsection{Improvements to GLOSS:}
\label{sec:improvements}

We asked participants for suggestions on improving GLOSS to better meet their needs. One common idea was to allow for query history and follow-up questions. P3 mentioned, ``\textit{If the response I got didn’t align exactly with what I was looking for, it’d be cool to follow up on it.}'' Meanwhile, P2 suggested, ``\textit{Maybe you could add a way to toggle through previous queries to see what those were.}'' Participants also felt that having instructions on how to prompt GLOSS for the best results would be helpful. P2 expressed, ``\textit{The only thing I worry about is whether I have the correct phrasing for the query so the model knows what I’m trying to ask. A cheat sheet for prompting, especially for people less familiar with language models, would be useful.}'' This reflects recent research showing that non-experts struggle with prompt design \cite{zamfirescu2023johnny}. Other suggested improvements included the ability to download code and data directly from the GLOSS interface. P1 also mentioned that having a feature to generate graphs in GLOSS would be very useful. 

These interviews serve as a strong indicator of GLOSS's potential impact in assisting sensemaking for non-CS researchers. In the future, we plan to conduct more comprehensive evaluations by deploying GLOSS in multiple research labs that work with passive sensing data.

\subsection{Narratives for Personal Health Reflection}

\begin{figure}[h]
  \centering
  \includegraphics[width=0.90\linewidth]{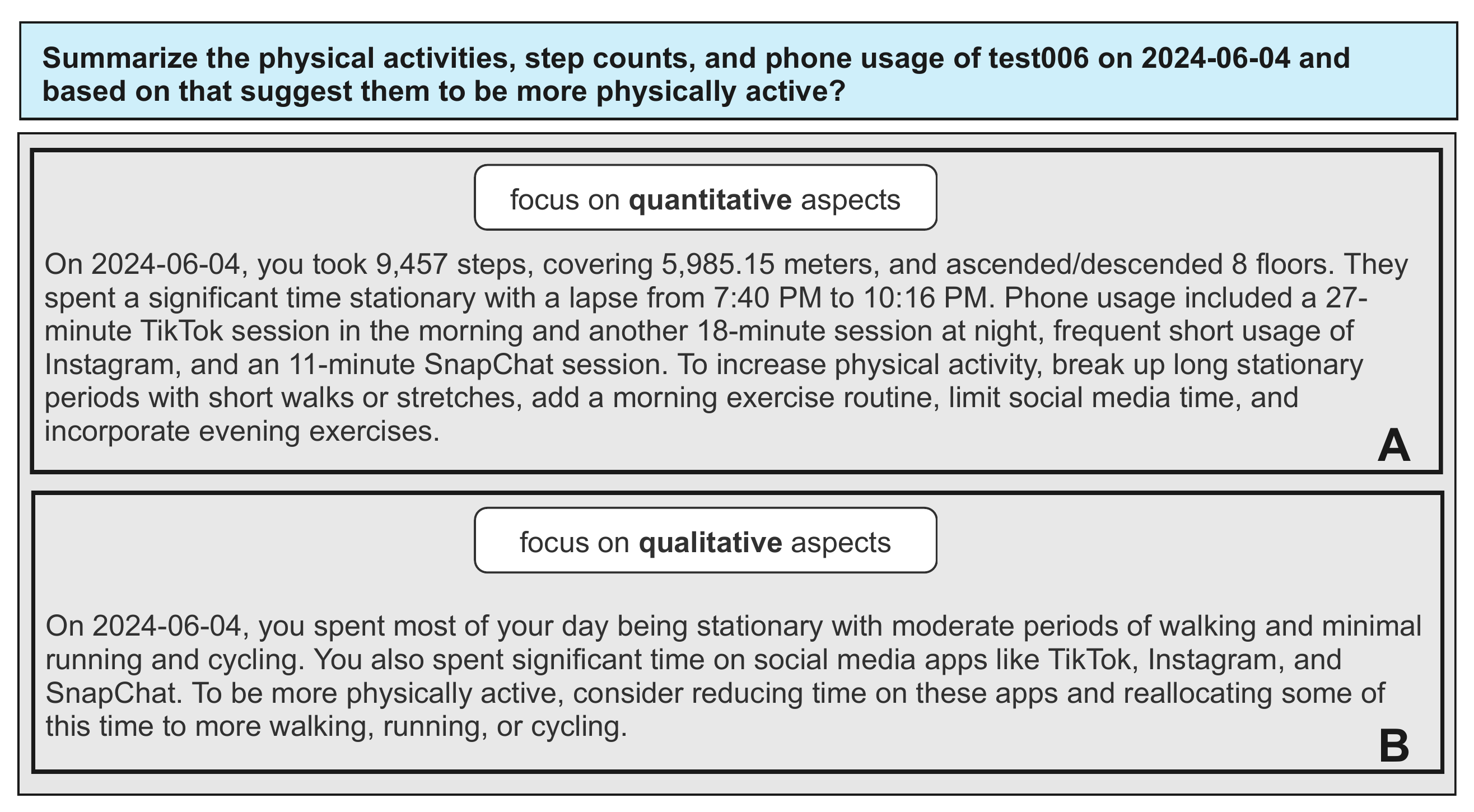}
  \caption{GLOSS generates different narratives to the same query when instructed to focus on quantitative and qualitative dimensions of wellbeing.}
  \label{fig:reflection}
\end{figure}

Reflection plays an important role in the journey of personal health informatics ~\cite{li2010stage}. It enables individuals to make meaningful connections with their health data \cite{li2011understanding} inspiring positive lifestyle changes ~\cite{botros2016go, diclemente2001role} and goal-setting behaviors ~\cite{niess2018supporting, gulotta2016fostering}. Merely presenting data, however, might not trigger reflection ~\cite{baumer2014reviewing}. Moreover, traditional ways of presenting data using tables and graphs often reduce individuals to quantified beings ~\cite{bentvelzen2022revisiting} overlooking more qualitative and holistic views of wellbeing. More recently, researchers used LLMs to generate personal health narratives that resonate more deeply with individuals' experiences and perceptions of wellbeing ~\cite{stromel2024narrating, fang2024physiollm, merrill2024transforming}. Stromel et al. \cite{stromel2024narrating}  demonstrated that supplementing step count visualizations with narrative descriptions significantly enhanced users' reflective engagement with their personal health data. GLOSS can be used to take this approach further by triangulating data from multiple streams and synthesizing multi-dimensional personalized narratives to support reflective practices. Furthermore, through presentation instructions, GLOSS can tailor data narratives to be adjusted to an individual’s needs. In Figure \ref{fig:reflection}, we show that using the same query but with different presentation instructions, GLOSS can generate narratives specifically tailored to different aspects of well-being (qualitative vs quantitative).

\subsection{AI-triggered Prompting}
Ecological momentary assessments (EMA) ~\cite{shiffman2008ecological} have been used extensively to collect various subjective measures of health and wellbeing, \rnr{such as, depression~\cite{xu2022globem, wang2014studentlife}, anxiety~\cite{tlachac_deprest-cat_2022}, daily activities~\cite{lecollecting2024, le_feasibility_2025}, and others \cite{campbell2023patient}}.
Traditionally, EMA prompts have been delivered at fixed times during the day. Recent works, however,  have leveraged AI models to identify opportune moments for delivering these prompts, a method known as AI-triggered prompting ~\cite{howe2022design, liao2018just, mishra2021detecting}. For instance, researchers have used AI models to predict physiological stress and use it as a basis for triggering EMA prompts aimed at capturing stressors ~\cite{neupane2024momentary}. Understanding the nature of stressors can help researchers design tailored and effective interventions. 

\begin{figure}[h]
  \centering
  \includegraphics[width=0.85\linewidth]{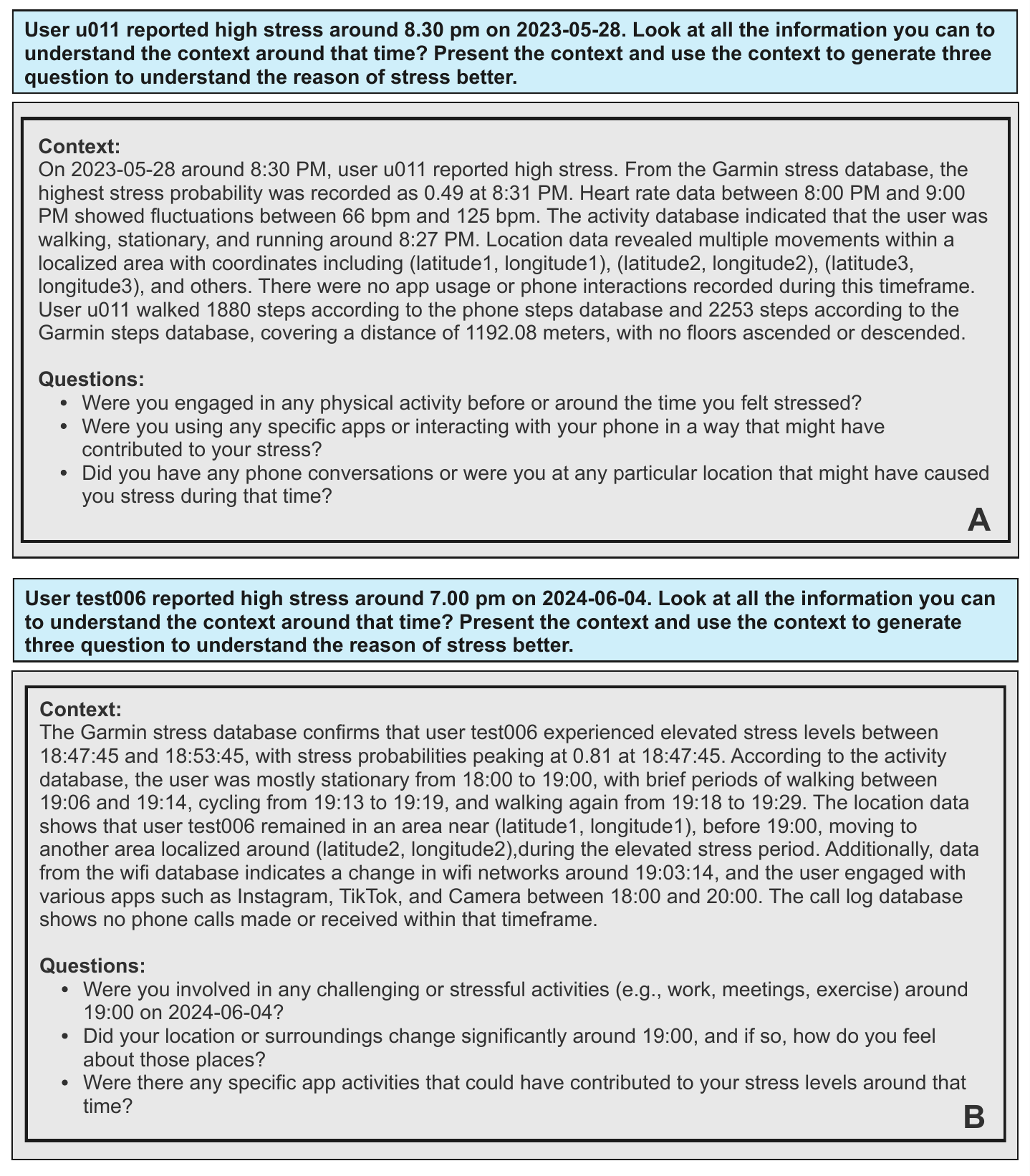}
  \caption{GLOSS can model the context using passive sensing data and then use the context to generate relevant questions to understand the stressor}
  \label{fig:stressor}
\end{figure}

While EMAs offer valuable insights into perceived subjective well-being, participants often find them burdensome~\cite{huang2015insufficient, campbell2023patient, choube2024sesame}, with the level of burden increasing as the length of the questionnaire increases~\cite{vhaduri2017design}. We demonstrate a use case for GLOSS where it can reduce the length of EMAs by only prompting contextually relevant questions. GLOSS can run in the background and keep integrating data from multiple sensor streams to continuously analyze and understand an individual's context. This context-aware understanding can be used to trigger only the prompts that are relevant to the context. For instance, in Figure \ref{fig:stressor} B, due to a change in location and connected Wi-Fi networks, GLOSS posed a question asking whether the change in surroundings was a potential cause of stress. \rnr{Moreover, it can also help in addressing missing data in passive sensing through relevant questions. For example, in Figure \ref{fig:stressor} A, due to the absence of app usage and phone conversation data—likely caused by technical issues—GLOSS proactively poses a question to address this data gap.}

\subsection{Enhancing Interpretability}

\begin{figure}[h]
  \centering
  \includegraphics[width=0.90\linewidth]{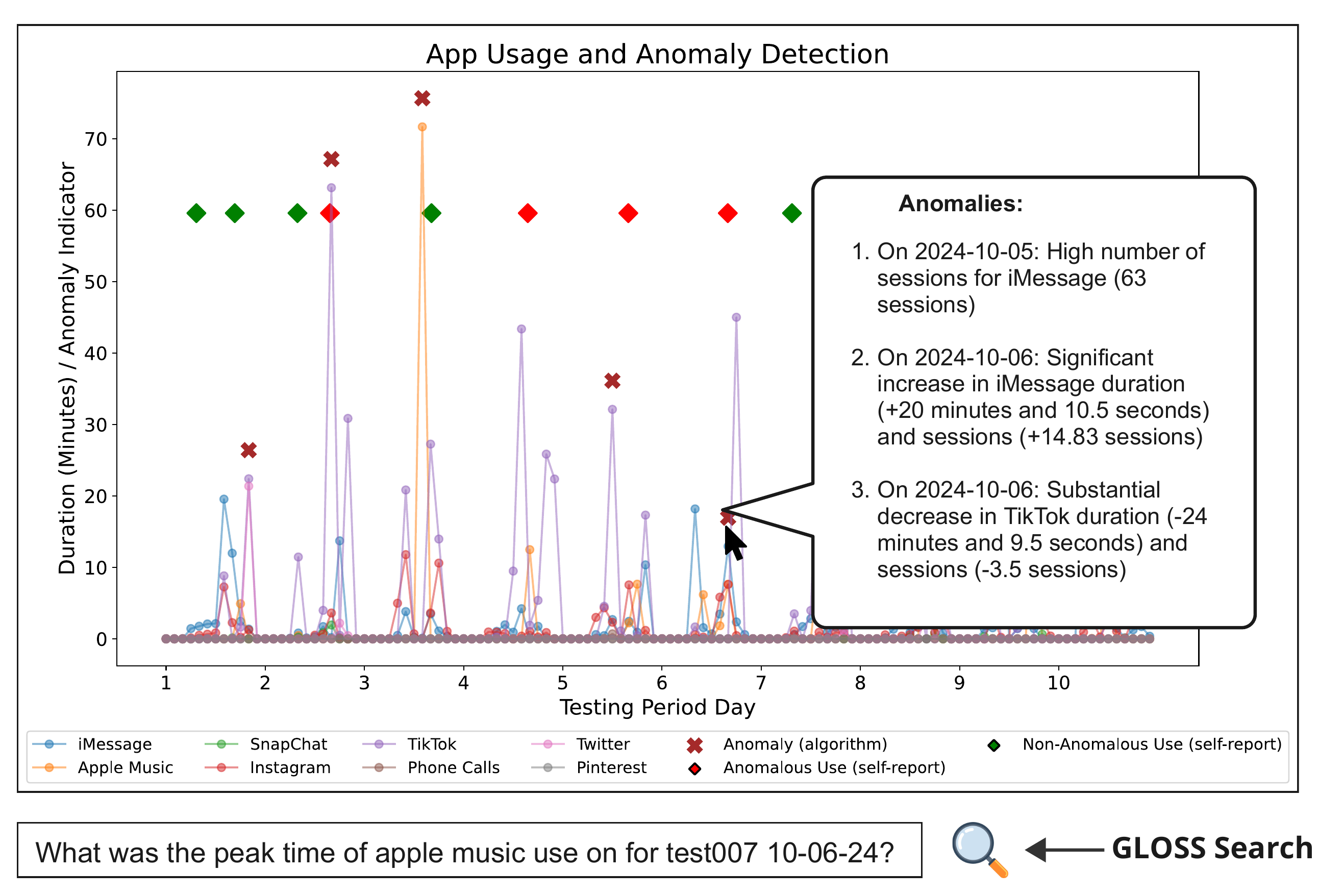}
  \caption{Black box models to be augmented with GLOSS generated insights to enhance interpretability of results supporting actionable steps for researchers. }
  \label{fig:anomaly}
\end{figure}

Deep learning models are extensively used to detect anomalous social media and phone use and its association with various mental health outcomes \cite{chancellor2020methods, DasSwain2022semantic, xu_leveraging_2021}. \rnr{Understanding these anomalous patterns can provide researchers with insights into how social media and phone use relate to various affective and mental health outcomes, ultimately aiding in the development of effective interventions \cite{wu2024mindshift}.} While \rnr{these anomaly detection models} demonstrate high accuracy, they are often considered black boxes, lacking interpretability \cite{zhang2021survey}. This limitation can hinder the ability to derive actionable steps or interventions \cite{khan2023drawbacks, durstewitz2019deep}. To address this issue, we present an example of how GLOSS can be integrated with traditional models to enhance interpretability and support researchers in taking actionable steps. In an ongoing project focused on detecting anomalous app usage, we implemented an unsupervised outlier detection model based on Long Short-Term Memory (LSTM) networks. These personalized models analyzed the past seven days of data to identify whether app usage on a given day was anomalous. As LSTMs are deep learning-based models, their predictions lacked interpretability. To overcome this, we augmented model predictions with insights derived from GLOSS (Figure \ref{fig:anomaly}), enhancing the interpretability of the results. Researchers can leverage these insights as a starting point and can query GLOSS for additional information to formulate actionable strategies.

\section{Discussion} 
In this section, we discuss the results from our evaluations of GLOSS as well as some implications and limitations of our work.

\subsection{GLOSS: Enabling Open-ended Sensemaking}
Researchers in the past have utilized passive sensing data to develop prediction and detection models for multiple but specific behavioral and health outcomes such as depression, stress, and physical activity \cite{xu2023globem, mishra_evaluating_2020, tlachac_deprest-cat_2022}. While these detection systems have shown tremendous potential, they often reduce complex aspects of health and well-being to binary predictions, failing to capture the broader, holistic picture \cite{bentvelzen2022revisiting}. There is a growing call within the research community to move beyond quantitative assessments and develop a more comprehensive understanding of well-being \cite{adler2024beyond, epstein_mapping_2020, stromel2024narrating}. To achieve this, we need systems that make passive sensing data more accessible and interpretable to a diverse range of stakeholders, including researchers, behavior scientists, clinicians, and even self-health trackers themselves. To this end, we designed a sensemaking system GLOSS (\textbf{G}roup of \textbf{L}LMs for \textbf{O}pen-ended \textbf{S}en\textbf{s}emaking) inspired by prior theories on human sensemaking \cite{klein2007data, pirolli2005sensemaking}. GLOSS is a query-based system that works with high-dimensional raw passive sensing data. GLOSS uses code generation techniques that are capable of writing and executing code to perform advanced data processing and analysis. Furthermore, GLOSS is able to triangulate multiple data streams to derive insights and understand the context of individuals. Using GLOSS does not demand any expertise in passive sensing or programming. Overall, GLOSS enables open-ended sensemaking, which was not possible with existing systems, paving the way for new applications and research directions.

\subsection{\rnr{Comparison of GLOSS and RAG Performance}}
Through our evaluations, we establish that GLOSS has promising accuracy (Table \ref{tab:model_comparison}), which is significantly higher than the commonly employed RAG techniques \cite{yang_retrieval-augmented_2025}. Our results align with prior research that shows RAG techniques perform poorly on mathematical or computation tasks \cite{agrawal_mindful-rag_2024, lewkowycz_solving_2022}. The consistency of GLOSS was affected by the ambiguity of queries, leading to GLOSS choosing accurate but different logical steps in different runs. \rnr{For instance, when answering the query "How many times did test008 use their phone during the commute on 2024-07-12?", GLOSS took different logical paths in separate runs: in one instance, it identified commute periods based on changes in location, while in another, it relied on activity recognition (walking or automotive) to determine commute times.} 
While different logical steps can encourage stakeholders to consider multiple perspectives on ambiguous queries, they can also create confusion. To mitigate this, we plan to introduce a preliminary step before sensemaking starts, where GLOSS can request more information to make the query more specific if it is deemed ambiguous. \rnr{In our example, we can ask the user to clearly define \textit{commute} or ask the user to choose one from possible options to calculate commute times (e.g., using \textit{activities} or using \textit{location}) to resolve the ambiguity. Another approach would be to expand GLOSS further to be comprehensive and combine several valid approaches to deal with potentially ambiguous tasks.}

In our subjective evaluations, GLOSS responses outperformed the RAG baseline in relevance, interpretation, domain knowledge, and logic, while RAG was rated higher in clarity. Based on the reasonings of our annotators, we believe GLOSS was over-explaining details as we set presentation instructions to \textit{``Explain clearly and in details"}. We did not do extensive experiments on presentation instructions affecting subjective evaluations of GLOSS responses. Although GLOSS outperformed RAG in most subjective dimensions, these differences were subtle, which is concerning as RAG responses were highly inaccurate. Our results are supported by prior works that have shown that LLMs can generate believable, confident and clear explanations for inaccurate results~\cite{turpin_language_2023, kunz_properties_2024}. These misleading explanations can be potentially detrimental, especially if applied in the health and well-being domain (e.g., for self-reflection, intervention design, etc.). \rnr{Thus, researchers should be cautious when using RAG for passive sensing tasks involving computation and data triangulation.}

\subsection{Potential for Future Applications}
We envision GLOSS as a foundational model for open-ended sensemaking, capable of being used directly or built upon for multiple use cases, including in human-facing or automated applications. In Section \ref{sec:usecase}, we demonstrated four applications of GLOSS inspired by ongoing and prior research in passive sensing. First, we conducted semi-structured interviews and a think-aloud session with three psychologists working with passive sensing data on a daily basis. The psychologists found GLOSS to be a highly valuable, time-saving, transparent, and easy-to-use system. They expressed that GLOSS enables them to perform sensemaking and analysis that was previously beyond their expertise. Second, we showed that GLOSS can be used to generate multi-dimensional narratives of personal health, which can be used to trigger reflective practices. Third, we showed GLOSS's application in generating relevant EMA prompts based on the context of an individual. Fourth, we showed how GLOSS can add to the interpretability of black box models, which can help in deriving actionable insights. These applications covered a diverse range of stakeholders, including psychologists (non-CS researchers), intervention designers, self-health trackers, and researchers. Through these examples, we wanted to highlight the breadth of applications of GLOSS, keeping a thorough evaluation of these use cases as part of our future work. As GLOSS is a core component that can be integrated with many human-facing and backend systems, GLOSS has several other potential applications. For instance, researchers can integrate GLOSS with text-to-speech and speech-to-text systems to enhance the accessibility of personal health data to individuals with vision-related disabilities and older adults who find visualization-only interfaces inaccessible. Furthermore, as the design of GLOSS does not restrict it to passive sensing, it can be extended for sensemaking in other forms of data (e.g., images and audio). Overall, we believe that GLOSS is a valuable sensemaking system for the UbiComp and HCI communities, and by making our code and results publicly available, we hope to support other researchers in exploring these possibilities with GLOSS.

\subsection{Future Improvements to GLOSS}
Based on our evaluations, we identified opportunities for improvements that can further enhance sensemaking using GLOSS. Some of these improvements were highlighted by participants in Section \ref{sec:improvements}. First, GLOSS is currently a single-turn query-based system. It does not keep a history of previous queries or does not allow users to ask follow-up questions. This hinders users from iteratively using GLOSS to understand data building on previous GLOSS responses. In the future, we plan to enable multi-turn conversations and keep a history of queries and results. Second, GLOSS UI currently does not support showing graphs or downloading data and code; we can add these functionalities to enhance the interpretability and transparency of results, and broaden the scope of use for researchers interested in using GLOSS for their studies. Third, GLOSS is sensitive to ambiguous queries, and some non-CS researchers have expressed concern about whether they are providing the correct prompts to the model. As non-experts often face trouble with prompt designing \cite{zamfirescu2023johnny}, we can include instructions on prompting with some examples in GLOSS UI to help users design effective prompts. Furthermore, with multi-turn conversation, GLOSS could request the user to clarify the query further in case it was unclear. These improvements can make GLOSS an even more powerful and valuable system.

\subsection{Limitations}
While our work offers a novel and valuable system for open-ended sensemaking, it has some limitations. 
First, we did not experiment with using multiple LLMs. As different LLMs have different code generation capabilities, our results may vary across different LLMs. 
Second, As GLOSS calls the GPT-4o model multiple times, its latency depends on the latency of these API calls and network connectivity. Thus, sometimes GLOSS has a latency of a couple of minutes when it processes complex queries, which can be irritating to users in applications where they expect near real-time latency. Having a local LLM in the system running GLOSS can reduce latency significantly.   
Third, we did not experiment with modifying presentation instructions and its impact on the subjective evaluation of user queries. Modifying presentation instructions can change the answer and thus their subjective evaluation on dimensions like \textit{clarity} and \textit{interpretation}. 
Fourth, we used passive sensing data modalities from phones and wearables. While GLOSS functioning does not rely on a single type of data, we still might need more evaluations to establish GLOSS's performance on other kinds of passive sensing data (for instance, ambient temperature sensors, audio devices, etc.).
Lastly, our query dataset for evaluation was small and was generated by researchers working in the passive sensing field. \rnr{Their queries naturally leaned more toward the objective inquiries they often encountered in their research, leading to more objective queries than subjective queries in our evaluation dataset. As future work, we will evaluate the effectiveness of GLOSS in diverse populations and query distributions. }

\section{Conclusion}

In the work, we presented a query-based novel sensemaking system  GLOSS where LLM agents interact according to theories on human sensemaking. GLOSS fills the gaps in existing sensemaking systems by enabling accurate, open-ended, and multimodal understanding of passive sensing data. We presented four use cases of GLOSS for automated tasks and human-facing tasks. The human subjects highly liked GLOSS, deeming it a valuable system that can save significant time and effort. We firmly believe GLOSS can be used to drive several interesting future health and well-being applications.





\ifanonymized
 \relax
\else
\fi 


\bibliographystyle{plain}	

\bibliography{references}

\end{document}